\documentstyle[12pt]{article}

  \def\pp{{\mathchoice
          {%displaystyle
              \kern 1pt%
              \raise 1pt
              \vbox{\hrule width5pt height0.4pt depth0pt
                    \kern -2pt
                    \hbox{\kern 2.3pt
                          \vrule width0.4pt height6pt depth0pt
                          }
                    \kern -2pt
                    \hrule width5pt height0.4pt depth0pt}%
                    \kern 1pt
           }
            {%textstyle
              \kern 1pt%
              \raise 1pt
              \vbox{\hrule width4.3pt height0.4pt depth0pt
                    \kern -1.8pt
                    \hbox{\kern 1.95pt
                          \vrule width0.4pt height5.4pt depth0pt
                          }
                    \kern -1.8pt
                    \hrule width4.3pt height0.4pt depth0pt}%
                    \kern 1pt
            }
            {%scriptstyle
              \kern 0.5pt%
              \raise 1pt
              \vbox{\hrule width4.0pt height0.3pt depth0pt
                    \kern -1.9pt  %[e]=0.15pt
                    \hbox{\kern 1.85pt
                          \vrule width0.3pt height5.7pt depth0pt
                          }
                    \kern -1.9pt
                    \hrule width4.0pt height0.3pt depth0pt}%
                    \kern 0.5pt
            }
            {%scriptscriptstyle
              \kern 0.5pt%
              \raise 1pt
              \vbox{\hrule width3.6pt height0.3pt depth0pt
                    \kern -1.5pt
                    \hbox{\kern 1.65pt
                          \vrule width0.3pt height4.5pt depth0pt
                          }
                    \kern -1.5pt
                    \hrule width3.6pt height0.3pt depth0pt}%
                    \kern 0.5pt%}
            }
        }}

  \def\mm{{\mathchoice
                       {%displaystyle
                             \kern 1pt
               \raise 1pt    \vbox{\hrule width5pt height0.4pt depth0pt
                                  \kern 2pt
                                  \hrule width5pt height0.4pt depth0pt}
                             \kern 1pt}
                       {%textstyle
                            \kern 1pt
               \raise 1pt \vbox{\hrule width4.3pt height0.4pt depth0pt
                                  \kern 1.8pt
                                  \hrule width4.3pt height0.4pt depth0pt}
                             \kern 1pt}
                       {%scriptstyle
                            \kern 0.5pt
               \raise 1pt
                            \vbox{\hrule width4.0pt height0.3pt depth0pt
                                  \kern 1.9pt
                                  \hrule width4.0pt height0.3pt depth0pt}
                            \kern 1pt}
                       {%scriptscriptstyle
                           \kern 0.5pt
             \raise 1pt  \vbox{\hrule width3.6pt height0.3pt depth0pt
                                  \kern 1.5pt
                                  \hrule width3.6pt height0.3pt depth0pt}
                           \kern 0.5pt}
                       }}

\catcode`@=11
\def\un#1{\relax\ifmmode\@@underline#1\else
        $\@@underline{\hbox{#1}}$\relax\fi}
\catcode`@=12

\let\du=\du                     % dot-under
\def\a{\alpha}
\def\b{\beta}
\def\c{\chi}
\def\d{\delta}
\def\e{\epsilon}
\def\f{\phi}
\def\g{\gamma}
\def\h{\eta}

\def\j{\psi}

\def\l{\lambda}
\def\m{\mu}
\def\n{\nu}
\def\o{\omega}

\def\q{\theta}
\def\r{\rho}
\def\s{\sigma}

\def\F{\Phi}
\def\G{\Gamma}

\def\L{\Lambda}
\def\O{\Omega}
\def\P{\Pi}

\def\U{\Upsilon}
\def\X{\Xi}

\def\ve{\varepsilon}

\def\cc{{\cal C}}

\def\ce{{\cal E}}

\def\ch{{\cal H}}

\def\cm{{\cal M}}

\def\cx{{\cal X}}
\def\cy{{\cal Y}}

\def\bo{{\raise-.5ex\hbox{\large$\Box$}}}               % D'Alembertian
\def\pa{\partial}                                       % curly d
\def\de{\nabla}                                         % del
\def\pr{\prod}                                          % product
\def\TH{{\raise.2ex\hbox{$\displaystyle \bigodot$}\mskip-4.7mu \llap H \;}}
\def\face{{\raise.2ex\hbox{$\displaystyle \bigodot$}\mskip-2.2mu \llap {$\ddot
        \smile$}}}                                      % happy face
\def\dg{\sp\dagger}                                     % hermitian conjugate
\def\sp#1{{}^{#1}}                              % superscript (unaligned)
\def\Bar#1{\overline{#1}}                       % big bar
\def\leftrightarrowfill{$\mathsurround=0pt \mathord\leftarrow \mkern-6mu
        \cleaders\hbox{$\mkern-2mu \mathord- \mkern-2mu$}\hfill
        \mkern-6mu \mathord\rightarrow$}
\def\dvec#1{\vbox{\ialign{##\crcr
        \leftrightarrowfill\crcr\noalign{\kern-1pt\nointerlineskip}
        $\hfil\displaystyle{#1}\hfil$\crcr}}}           % <--> accent
\def\dt#1{{\buildrel {\hbox{\LARGE .}} \over {#1}}}     % dot-over for sp/sb
\def\frac#1#2{{\textstyle{#1\over\vphantom2\smash{\raise.20ex
        \hbox{$\scriptstyle{#2}$}}}}}                   % fraction
\def\sfrac#1#2{{\vphantom1\smash{\lower.5ex\hbox{\small$#1$}}\over
        \vphantom1\smash{\raise.4ex\hbox{\small$#2$}}}} % alternate fraction
\def\bfrac#1#2{{\vphantom1\smash{\lower.5ex\hbox{$#1$}}\over
        \vphantom1\smash{\raise.3ex\hbox{$#2$}}}}       % "
\def\afrac#1#2{{\vphantom1\smash{\lower.5ex\hbox{$#1$}}\over#2}}    % "
\def\[{\lfloor{\hskip 0.35pt}\!\!\!\lceil}
\def\]{\rfloor{\hskip 0.35pt}\!\!\!\rceil}

\def\du#1#2{_{#1}{}^{#2}}
\def\ud#1#2{^{#1}{}_{#2}}

\def\fracm#1#2{\hbox{\large{${\frac{{#1}}{{#2}}}$}}}
\def\ha{{\fracmm12}}

\def\un{\underline}
\def\fracmm#1#2{{{#1}\over{#2}}}

\def\low#1{{\raise -5pt\hbox{${\hskip 0.75pt}\!_{#1}$}}}

\def\Dot#1{\buildrel{_{_{\hskip 0.01in}\bullet}}\over{#1}}
\def\dt#1{\Dot{#1}}

\newskip\humongous \humongous=0pt plus 1000pt minus 1000pt
\def\caja{\mathsurround=0pt}
\def\eqalign#1{\,\vcenter{\openup2\jot \caja
        \ialign{\strut \hfil$\displaystyle{##}$&$
        \displaystyle{{}##}$\hfil\crcr#1\crcr}}\,}
\newif\ifdtup

\def\ref#1{$\sp{#1)}$}

\def\pl#1#2#3{Phys.~Lett.~{\bf {#1}B} (19{#2}) #3}
\def\np#1#2#3{Nucl.~Phys.~{\bf B{#1}} (19{#2}) #3}

\def\pr#1#2#3{Phys.~Rev.~{\bf D{#1}} (19{#2}) #3}
\def\cqg#1#2#3{Class.~and Quantum Grav.~{\bf {#1}} (19{#2}) #3}

\def\ibid#1#2#3{{\it ibid.}~{\bf {#1}} (19{#2}) #3}

\parskip=\medskipamount                 % space between paragraphs (LaTeX)
\thicklines                         % thick straight lines for pictures (LaTeX)

\def\plpl{\raise-2pt\hbox{$\raise3pt\hbox{$_+$}\hskip-6.67pt\raise0.0pt}}
\def\mimi{\raise-2pt\hbox{$\raise3pt\hbox{$_-$}\hskip-6.67pt\raise0.0pt}}

\def\dvm{\raisebox{-.145ex}{\rlap{$=$}}}
\def\DM{{\scriptsize{\dvm}}~~}
\def\lin{\vrule width0.5pt height5pt depth1pt}
\def\dpx{{{ =\hskip-3.75pt{\lin}}\hskip3.75pt }}

\def\beq{\begin{equation}}
\def\eeq{\end{equation}}

\def\beqx{\begin{displaymath}}
\def\eeqx{\end{displaymath}}

\def\beql{\begin{eqnarray}}
\def\eeql{\end{eqnarray}}

\newskip\humongous \humongous=0pt plus 1000pt minus 1000pt
\def\caja{\mathsurround=0pt}

\begin{document}

% =========================== title page ==================================

\thispagestyle{empty}               % no heading or foot on title page (LaTeX)

\def\border{                                            % border
        \setlength{\unitlength}{1mm}
        \newcount\xco
        \newcount\yco
        \xco=-24
        \yco=12
        \begin{picture}(140,0)
        \put(-20,11){\tiny Institut f\"ur Theoretische Physik Universit\"at
Hannover~~ Institut f\"ur Theoretische Physik Universit\"at Hannover~~
Institut f\"ur Theoretische Physik Hannover}
        \put(-20,-241.5){\tiny Institut f\"ur Theoretische Physik Universit\"at
Hannover~~ Institut f\"ur Theoretische Physik Universit\"at Hannover~~
Institut f\"ur Theoretische Physik Hannover}
        \end{picture}
        \par\vskip-8mm}

\def\headpic{                                           % UH heading
        \indent
        \setlength{\unitlength}{.8mm}
        \thinlines
        \par
        \begin{picture}(29,16)
        \put(75,16){\line(1,0){4}}
        \put(80,16){\line(1,0){4}}
        \put(85,16){\line(1,0){4}}
        \put(92,16){\line(1,0){4}}

        \put(85,0){\line(1,0){4}}
        \put(89,8){\line(1,0){3}}
        \put(92,0){\line(1,0){4}}

        \put(85,0){\line(0,1){16}}
        \put(96,0){\line(0,1){16}}
        \put(79,0){\line(0,1){16}}
        \put(80,0){\line(0,1){16}}
        \put(89,0){\line(0,1){16}}
        \put(92,0){\line(0,1){16}}
        \put(79,16){\oval(8,32)[bl]}
        \put(80,16){\oval(8,32)[br]}

        \end{picture}
        \par\vskip-6.5mm
        \thicklines}

\border\headpic {\hbox to\hsize{
\vbox{\noindent DESY~ 96 --165 \hfill  hep-th/9608131  \\
                ITP--UH--15/96 \hfill revised version  }}}

\noindent
\vskip1.3cm
\begin{center}
{\Large\bf (4,4) SUPERFIELD SUPERGRAVITY~\footnote{Supported in part by 
the `Deutsche Forschungsgemeinschaft' and the NATO grant CRG 930789}}\\
\vglue.3in

Sergei V. Ketov, \footnote{
On leave of absence from:
High Current Electronics Institute of the Russian Academy of Sciences,
\newline ${~~~~~}$ Siberian Branch, Akademichesky~4, Tomsk 634055, Russia} and
Christine Unkmeir

{\it Institut f\"ur Theoretische Physik, Universit\"at Hannover}\\
{\it Appelstra\ss{}e 2, 30167 Hannover, Germany}\\
{\sl ketov@itp.uni-hannover.de}

and

Sven-Olaf Moch  

{\it Deutsches Elektronen Synchrotron, DESY}\\
{\it Notkestra\ss{}e 85, 22603 Hamburg, Germany}\\
{\sl moch@mail.desy.de}
\end{center}
\vglue.2in
\begin{center}
{\Large\bf Abstract}
\end{center}
We present the N=4 superspace constraints for the two-dimensional (2d) 
off-shell (4,4) supergravity with the superfield strengths expressed in terms 
of a (4,4) twisted (scalar) multiplet TM-I, as well as the corresponding 
component results, in a form suitable for applications. The constraints are shown
to be invariant under the N=4 super-Weyl transformations, whose N=4 superfield
parameters form another twisted (scalar) multiplet TM-II. To solve the 
constraints, we propose the Ansatz which makes the N=4 superconformal flatness 
of the N=4 supergravity curved superspace manifest. The locally (4,4) 
supersymmetric TM-I matter couplings, with the potential terms resulting from 
spontaneous supersymmetry breaking, are constructed. We also find the full (4,4) 
superconformally invariant (improved) TM-II matter action. The latter can be 
extended to the (4,4) locally supersymmetric Liouville action which is suitable 
for describing (4,4) supersymmetric non-critical strings.

\newpage

\section{Introduction}

A full off-shell structure of any supersymmetric field theory most naturally 
exhibits itself in superspace, provided the superfield formulation of the 
theory in terms of unconstrained superfields (the so-called 
{\it prepotentials}) is available. This is particularly relevant for the 
supergravity theories, which are usually formulated in superspace by using
the Wess-Zumino-type constraints~\cite{wz} (see refs.~\cite{bw,book1,book2} for
a review.) A fully covariant superfield formulation is desirable for 
quantisation purposes, as well as for renormalisation or a finiteness check. 
A covariant superspace solution is also useful for studies of super-Riemannian 
surfaces and the associated super-Beltrami differentials, where conformal 
gauge may not be convenient and light-cone gauge may not be accessible, e.g.
as far as the higher-genus string and superstring amplitudes are 
concerned~\cite{kni}.

Once a full set of auxiliary fields needed to close the supersymmetry algebra
in a supersymmetric field theory is known, it should be possible to solve the 
equivalent superspace constraints. In four dimensions, the full solution to 
the N=1 superspace supergravity is known for a long time~\cite{ws}, whereas 
solving the N=2 
extended superspace supergravity presumably requires the use of the N=2 
harmonic superspace \cite{hrs}, with the necessarily infinite number of 
auxiliary fields. As far as the four-dimensional N=2 supergravity in the 
ordinary N=2 superspace is concerned, only linearised solutions were found 
so far~\cite{gs,ket}.

In {\it two dimensions} (2d), where the Lorentz group is more restricted, it 
should be possible to find full covariant solutions to the $(p,q)$-extended 
supergravities in the ordinary $(p,q)$-extended superspace, whenever the 
corresponding off-shell formulation is available, i.e. if $p,q\leq 4$. Indeed, 
the fully covariant solutions are already known for $(1,0)$ \cite{ggmt}, 
$(1,1)$ \cite{gn}, $(p,0)$ \cite{eo} and $(2,2)$ \cite{a} supergravities. In 
particular, the solution to the 2d, $(2,2)$ supergravity can also be obtained 
by dimensional reduction from {\it four dimensions} (4d). Though being not 
practical for solving superspace constraints, the method of dimensional 
reduction is nevertheless useful for getting insights into the complicated 
component structure of extended supergravities, and for spontaneous 
supersymmetry breaking as well (see sect.~4 for an example).

To the best of our knowledge, no attempts were ever made towards solving the 
covariant 2d off-shell (4,4) superspace supergravity constraints, since they 
were first formulated by Gates {\it et.~al.} in ref.~\cite{glo} (see also the 
related work~\cite{ghn89}). Recently, Grisaru and Wehlau \cite{gw1,gw2} found
the complete covariant solution to the 2d, (2,2) supergravity constraints in 
the ordinary N=2 superspace, as well as the corresponding superspace measures 
and invariant actions. It was achieved, in part, by working in a proper 
light-cone-type basis, rather then using the gamma matrices as in 
refs.~\cite{glo,ghn89}. In this paper, we begin the similar program for the 
case of the 2d, (4,4) superspace supergravity. Surprisingly enough, as far as the 
solution to the (4,4) supergravity constraints is concerned, it turns out to 
be possible to follow the lines of the N=2 solution up to a Wess-Zumino-type 
supersymmetric gauge fixing. The gauge-fixing should result in only one 
irreducible (4,4) superfield describing the off-shell N=4 conformal supergravity 
multiplet. It is related to the fact that the general (4,4) vector superfield 
$H^m$ has many redundant supersymmetric gauge degrees of freedom, unlike its N=2 
counterpart. The relevant irreducible superfield can be rather easily identified 
in the linearised approximation~\cite{kr}. Gauging away the rest of the N=4 
superfields does not introduce propagating ghosts, despite of a high degree of 
non-linearity. The supersymmetric gauge-fixing in the (4,4) superspace 
supergravity is however beyond the scope of this paper.

We also present here some interesting new features for 2d couplings of the 
twisted chiral matter multiplets, TM-I and TM-II, to the (4,4) supergravity. 
In particular, we show how to generate the potential terms via spontaneous 
N=4 supersymmetry breaking by dimensional reduction. This approach can be 
considered as the alternative to the global symmetry gauging in the (4,4) 
extended supergravity with matter, which usually leads to the (classical) scalar 
potentials unbounded from below~\cite{mu,ko}. The known 
exception is the {\it Wess-Zumino-Novikov-Witten-Liouville}-type (WZNWL-type) 
{\it non-linear sigma-model} (NLSM), which reduces to an $SU(2)\times U(1)$ WZNW 
model in the limit of vanishing Liouville-type interaction~\cite{ik,ikl,gi}. It 
is precisely the WZNW model whose symmetry gauging amounts to the coupling with 
the (4,4) supergravitational background in the superconfomal gauge~\cite{kpr}. It
is of interest to know the full covariant and explicitly supersymmetric form of 
that NLSM, and the (4,4) superspace supergravity provides the natural framework 
for that purpose.

Our paper is organized as follows: in sect.~2 the N=4 superspace geometry and
the N=4 superfield supergravity constraints are discussed. Sect.~3 is devoted 
to the component structure of the scalar multiplets TM-I and TM-II. In sect.~3
we briefly review the solution to the N=2 superfield supergravity constraints 
as presented in ref.~\cite{gw1}, which constitutes the pattern we are going to 
follow to solve the N=4 constraints in the next sect.~4. In sect.~5 we 
construct the (4,4) locally supersymmetric 2d NLSMs out of TM-I and TM-II matter.
In particular, we find the fully covariant (4,4) supersymmetric extension of the 
Liouville theory, and generate potential terms due to the spontaneous 
supersymmetry breaking. Our conclusions are summarized in sect.~6. A part of our 
notation and conventions, as well as some useful identities, are collected in 
Appendix A. The component structure of the 2d, (4,4) supergravity multiplet is 
reviewed in Appendix B. In Appendix C we describe the dimensional reduction of 
the 4d reduced chiral N=2 superfield down to two dimensions, which generates the
scalar potential leading to spontaneous supersymmetry breaking.
\vglue.2in

\section{N=4~ superspace geometry}

The 2d minimal off-shell (4,4) supergravity in N=4 superspace was first 
formulated in ref.~\cite{glo}, with the particular 2d, (4,4) hypermultiplet 
(TM-II) as a scale compensator. There is, in fact, the whole variety of the 
so-called {\it variant} representations for a 2d, (4,4) 
hypermultiplet~\cite{gk4}. The variant representations are inequivalent since 
there is no way to convert one of them into another while keeping the (4,4) 
supersymmetry. To distinguish between the different variant representations of
the 2d, (4,4) hypermultiplet, we use the classification adopted in 
ref.~\cite{gk4}. For our purposes in this paper, we only need the two variant 
off-shell hypermultiplets, TM-I and TM-II. Both have four propagating scalars, 
which are all singlets in TM-I, while they form one triplet and one singlet in 
TM-II, with respect to the $SU(2)$ internal symmetry group rotating the N=4 
supersymmetry charges~\cite{gk4}. The TM-II is preferable for its use as a 
(4,4) scale compensator, since it has only one scalar which can represent the 
usual Weyl transformation parameter. Still, there is no obvious reason against
the use of the TM-I as a (4,4) scale compensator, even though it has four 
physical scalars on equal footing. Since we are not interested in presenting here
all possible versions of the N=4 supergravity, we choose its particular version
whose superfield strengths form a (4,4) locally supersymmetric TM-I while the 
(4,4) scale compensator is given by a TM-II, as in ref.~\cite{glo}.

Flat $N=4$ superspace in two dimensions is parameterised by the 
coordinates~\footnote{See Appendix A for more about our notation.}
$$ z^A ~=~ ( x^{\dpx}, x^{\DM}, \q^{+ i}, \q^{- i}, \q^{\Dot +}_{~~i}, 
        \q^{\Dot -}_{~~i} )~,\eqno(2.1) $$
where  $x^{\dpx}$ and $x^{\DM}$ are two real bosonic (commuting) coordinates, 
$\q^{\pm i}$ and their complex conjugates $\q^{\Dot\pm}_{~~i}$ are complex 
fermionic (anticommuting) coordinates, $i=1,2$. The fermionic coordinates 
$\q^{\pm i}$ are spinors with respect to $SU(2)$. Their complex conjugates were
defined by
$$
(\q^{\pm i})^* ~\equiv~-~\q^{\Dot \pm}_{~~i}~~~,\quad \q^{\Dot\pm ~i}=\cc^{ij}
\q^{\Dot \pm}_{~~j}~~~,\eqno(2.2) 
$$
where the star denotes usual complex conjugation. The $SU(2)$ indices are usually
`canonically' contracted from the upper left to the lower right 
(the North-West/South-East rule), otherwise an extra sign arises.  
These indices are raised and lowered by $\cc^{i j}$ and 
$\cc_{i j}$, whose explicit form is given by $\cc^{i j} = i\ve^{i j}$ and 
$(\cc^{i j})^* = \cc_{i j}$. We prefer to work in a light-cone-type basis, 
rather than using the 2d gamma matrices 
({\it cf.\/} refs.~\cite{glo,ghn89,gw1}).

The spinorial covariant derivatives in the {\it flat} (4,4) superspace satisfy 
the algebra 
$$
\{ D\low{+ i}\, ,D_{\Dot{+} j} \} ~=~ i~ \cc_{i j}~ \pa_{\dpx}~,\qquad
\{ D\low{- i}\, ,D_{\Dot{-} j} \} ~=~ i~ \cc_{i j}~ \pa_{\DM}~,\eqno(2.3)
$$
while all other (anti)commutators vanish.

The local symmetries of the N=4 superfield supergravity comprise the N=4 
superspace general coordinate transformations, local Lorentz frame rotations 
and $SU(2)$ internal frame rotations. Therefore, the fully covariant 
derivatives   in the {\it curved} N=4 superspace should include the tangent 
space generators for all that symmetries, with the corresponding 
connections~\cite{book1,book2}. The superspace geometry of any supergravity 
theory is described by suitable 
constraints on the torsion and curvature for the spinorial covariant 
derivatives. As far as the (4,4) curved superspace is concerned, we define
$$ \de_A ~=~ E_A^M D_M + \O_A \cm + i~ \G_{A}\cdot \cy~,\eqno(2.4)$$
where the N=4 supervielbein $E_A^M$, the Lorentz generator $\cm$ with the 
Lorentz connection $\O_A$, and the $SU(2)$ generators $\cy_i{}^j$ with the 
$SU(2)$ connection  $(\G_{A})_j{}^i$ have been introduced. We sometimes use the 
dot product, $\G_{A}\cdot\cy\equiv(\G_{A})_j{}^i\cy_i{}^j$, in order to simplify 
our notation. The operators $\de_A$ change covariantly under all the local 
symmetry transformations by definition, i.e.  
$$\de_A'=e^{-\ch}\de_A e^{\ch}~,\qquad \ch=H^M\pa_M +H\cm 
+ iH_j{}^i\cy_i{}^j~,\eqno(2.5)$$
where $H^M$, $H$, and $iH_j{}^i$ are the infinitesimal superfield parameters 
for the N=4 superspace general coordinate, local Lorentz and $SU(2)$ 
transformations, respectively. 

We assume that the supervielbein is invertible, and identify the lowest-order 
component in the $\q$-expansion of the superfield $E^a_{\m}$ with the zweibein,
$\left. E^a_{\m}\right|=e^a_{\m}$. Similarly, 
$\left. E^{i\pm}_{\m}\right|=\j^{i\pm}_{\m}$ and 
$\left. \G\du{\m j}{i}\right|=A\du{\m j}{i}$ define the rest of the gauge 
fields for the 2d conformal (4,4) supergravity. The superfield torsion and 
curvature tensors are defined as usual, namely
$$ \[ \de_A , \de_B \} = T\du{A B}{C}\de_C + R_{A B} \cm + i~ F_{A B}\cdot\cy~.
\eqno(2.6)$$

The generators for the local Lorentz and $SU(2)$ frame transformations are
defined by their action on spinors or the spinorial derivatives,
$$ \[ \cm , \de\low{\pm i} \] ~=~ \pm \frac{1}{2} \de\low{\pm i}~,\quad
\[ \cm , \de_{\Dot{\pm}}^{~~i} \] ~=~ \pm \frac{1}{2} \de_{\Dot{\pm}}^{~~i}~,
$$
$$\eqalign{
\[ \cy_i^{~j} , \de\low{\pm k} \] ~=~& + \d_k^{~j} \de\low{\pm i} - 
                      \frac{1}{2} \d_i^{~j} \de\low{\pm k} ~,\cr
\[ \cy_i^{~j} , \de_{\Dot{\pm}}^{~~k} ] ~=~& 
      -  \d_i^{~k} \de_{\Dot{\pm}}^{~~j} +  
                      \frac{1}{2} \d_i^{~j} \de_{\Dot{\pm}}^{~~k} ~.\cr}
\eqno(2.7)$$

The supervielbein and superconnections define a highly reducible representation
of N=4 supersymmetry, and they have therefore to be restricted by covariant 
constraints~\cite{dra}.

The constraints defining the N=4 superfield supergravity are given
by ({\it cf.}\/ ref.~\cite{glo})
$$
\{ \de_{\pm i}\, , \de_{\pm j} \} ~=~ 0~,\quad
\{ \de\low{+ i} \, , \de_{\Dot{+} j} \} ~=~ i \cc\low{i j} \de_{\dpx}~,\quad
\{ \de\low{- i} \, , \de_{\Dot{-} j} \} ~=~ i \cc\low{i j} \de_{\DM}~,$$
$$\{ \de_{+ i} \, , \de_-^{~~j}  \} ~=~ - \frac{i}{2}~R^*~
       \left( \d_i^{~j}\cm - \cy_i^{~j} \right) ~,\eqno(2.8)$$
$$ \{ \de\low{+ i}\, , \de_{\Dot -}^{~~j}  \} ~=~
      - \frac{i}{2}~ S~ \left( \d_i^{~j}\cm - \cy_i^{~j} \right) -
    \frac{1}{2}~ T~ \left( \d_i^{~j}\cm - \cy_i^{~j} \right) ~,$$
as well as their complex conjugates. Given the constraints above, the 
additional constraints on the (4,4) supergravity superfield strengths $R_{AB}$
and $(F_{AB})\du{j}{i}$ follow from the Bianchi identities. For instance, 
the Bianchi identity for the torsion $T^A= \de E^A$ reads  
$$ \de T^A = E^B~R_B^{~A}~.\eqno(2.9)$$
As far as the full algebra of the covariant derivatives is concerned, we find
\begin{eqnarray}
    \[ \de_{+ i} , \de_{\DM} \] &=&
 \fracmm{1}{4}~\Bigl[
~- 2 R^*~\de_{{\Dot -} i} + 2 (~S - iT~)~\de_{-i} \nonumber \\
& & + \left( \de_{\Dot-}^{~~~j} R^* -
   \de_{-}^{~j}~(~S - iT~)~\right)
  ~\left( \cc_{i j} \cm - \cy_{i j} \right)~\Bigr]~, \nonumber
\end{eqnarray}
\begin{eqnarray}
\[ \de_{- i} , \de_{\dpx} \] &=&
\fracmm{1}{4}~\Bigl[~- 2 R^* ~\de_{{\Dot +} i}
    - 2 (~S + iT~)~\de_{+i} \nonumber \\ 
& & - \left( \de_{\Dot+}^{~~~j}  R^* -
       \de_{+}^{~j}~(~S + iT~)~\right)
~\left( \cc_{ij} \cm + \cy_{ij} \right)~\Bigr]~, \nonumber
\end{eqnarray}
$$
\[ \de_{+ i} , \de_{\dpx} \] ~=~ 0~,\qquad
\[ \de_{- i} , \de_{\DM} \] ~=~ 0~,
$$
\begin{eqnarray}
\[ \de_{\dpx} , \de_{\DM} \] &=& 
\fracmm{i}{4}~ \Bigl[ ~ 
+ ( \de_{-}^{~~i}{R} ) \de_{+i} 
- ( \de_{+}^{~~i}{R} ) \de_{-i} 
- ( \de_{\Dot-}^{~~i} R^* ) \de_{\Dot+ i} 
+ ( \de_{\Dot+}^{~~i} R^* ) \de_{\Dot- i}  \nonumber \\ 
&&~~~~~~~~
- ( \de_{\Dot-}^{~~i}~ (S+iT) )  \de_{+i} 
+ ( \de_{-}^{~~i}~ (S- iT) )  \de_{\Dot+ i} \nonumber \\ 
&&~~~~~~~~
- ( \de_{\Dot+}^{~~i} ~ (S- iT) )  \de_{-i}
+ ( \de_{+}^{~~i}~ (S+iT) )  \de_{\Dot- i} 
~\Bigr]         \nonumber \\ 
&&
+\fracmm{1}{2}~ \Bigl[ ~
  R R^* - S^2 - T^2 
-\frac{i}{4}  ( \de_{+}^{~~i} \de_{-i} R ) 
+\frac{i}{4}  ( \de_{\Dot+}^{~~i} \de_{\Dot- i} R^* )  \nonumber \\ 
&&~~~~~~~~
-\frac{i}{4}  ( \de_{\Dot+}^{~~i} \de_{-i}~ (S- iT) )
+\frac{i}{4}  ( \de_{+}^{~~i} \de_{\Dot- i}~  (S+iT) ) 
~\Bigr] ~  \cm   \nonumber \\ 
&&
+\fracmm{i}{8}   \Bigl[ ~
 ( \de_{+}^{~~i} \de_{-j} R ) 
-  ( \de_{\Dot+}^{~~i} \de_{\Dot- j} R^* )  \nonumber \\
&&~~~~~~~~
+ ( \de_{\Dot+}^{~~i} \de\low{-j}~ (S- iT) )
-  ( \de\low{+}^{~~i} \de_{\Dot- j}~  (S+iT) )
~\Bigr] ~\cy_i^{~j}
 ~ ,   \nonumber   
\end{eqnarray}
$$
\[ \de_{\dpx} , \de_{\dpx} \] ~=~ 0~,\qquad
\[ \de_{\DM} , \de_{\DM} \] ~=~ 0~. \eqno(2.10)$$

The constraints following from the Bianchi identity 
$$\de F = 0 \eqno(2.11)$$ 
for the $SU(2)$ superfield strength $F$ are given by
$$\eqalign{
\de_{+k}~ F_{+i~-j} + \de_{+j}~ F_{+k~-i} ~= & 0~,\cr
\de_{-k}~ F_{+i~-j} + \de_{-i}~ F_{+j~-k} ~= & 0~,\cr
\de\low{+k}~ F_{+i~{\Dot{-}}j} + \de\low{+j}~ F_{+k~{\Dot{-}}i} ~= & 0~,\cr 
\de\low{-k}~ F_{{\Dot{+}}i~-j} + \de\low{-i}~ F_{{\Dot{+}}j~-k} ~= & 0~.\cr}
\eqno(2.12)$$

The defining constraints (2.8) also imply further consistency relations having 
the form
$$\eqalign{
 \de_{{\Dot{-}} k}~ F\low{+i~ -j} + \de\low{-i}~F_{+j~{\Dot{-}}k}
        - T\du{-i~ {\Dot -}k}{\DM}~ F\low{\DM +j}   ~=~& 0~, \cr
\de_{{\Dot{+}} k}~ F\low{+i~ -j} + \de\low{+j}~F_{\Dot{+}k~-i}
        - T\du{+j~ \Dot{+} k}{\dpx}~ F\low{\dpx -i}   ~=~& 0~.\cr}
\eqno(2.13)$$

Taken together, they lead to the certain constraints on the (4,4) supergravity 
field strengths which comprise the complex scalar superfield $R$ and the two real 
ones, $S$ and $T$. We find 
$$ \de_{\Dot{\pm} i} {R} ~=~0~,\quad
\de\low{{\pm} i}  {R} ~=~ \pm 2~ \de_{\Dot{\pm} i} S~,\quad
\de\low{{\pm} i} S ~=~ \pm ~i~ \de\low{{\pm} i} T~, $$
$$ \[ \de_+^{~~i},\de_{+i} \] R = \[ \de_-^{~~i},\de_{-i} \] R =0~, \eqno(2.14)  $$
and their conjugates, where the signs are correlated, as well as the additional 
reality condition 
$$ \left( \de\low{+i}\de_{\Dot{-}j}S\right)^* =
\de_+^{~~i}\de_{\Dot{-}}^{~~j}S~,\eqno(2.15)$$
Eqs.~(2.14) and (2.15) define the {\it twisted}-I hypermultiplet (TM-I). It is 
not difficult to check that the TM-I has $8_{\bf B}\oplus 8_{\bf F}$ 
independent off-shell degrees of freedom (see also the next sect.~3).

Some of the constraints given above were also found in ref.~\cite{glo}. In 
particular, it is straightforward to verify that no more consistency relations 
follow from the Bianchi identities. The (4,4) supergravity multiplet, 
comprising a graviton $e^a_{\m}$, four gravitini $\j_{\m}^{i\pm}$, an $SU(2)$ 
triplet of graviphotons $A_{\m}^I$, $I=1,2,3$, a complex scalar $R$, and two 
real scalars $S$ and $T$, appears at the component level. The supersymmetry 
transformation laws for the components of the (4,4) conformal supergravity 
multiplet are collected in Appendix B.
\vglue.2in

\section{TM-I and TM-II in curved (4,4) superspace}

In this section we provide the superspace formulation for two off-shell (4,4) 
hypermultiplets, TM-I and TM-II, in the presence of the (4,4) supergravity. The
rigid (4,4) supersymmetry hypermultiplets are known for a long time (see, e.g.,
ref.~\cite{gk4} for a recent review). Their minimal versions, TM-I and TM-II, 
each have  $8_{\bf B}\oplus 8_{\bf F}$ independent components.

The constraints defining the TM-I were already given in the preceding sect.~2,
namely,
$$ \de_{\Dot{\pm} i} {B} ~=~0~,\quad
\de\low{{\pm} i}  {B} ~=~ \pm 2~ \de_{\Dot{\pm} i}F ~,\quad
\de\low{{\pm} i} F ~=~ \pm ~i~ \de\low{{\pm} i} G~, $$
$$ \[ \de_{+}^{~~i},\de_{+ i} \] B = \[ \de_-^{~~i},\de_{-i} \] B =0~, \eqno(3.1)  $$
and their conjugates, in terms of four scalar superfields, the complex one $B$
and two real ones $F$ and $G$, with the additional reality condition
$$ \left( \de\low{+i}\de_{\Dot{-}j}F\right)^* =
\de_+^{~~i}\de_{\Dot{-}}^{~~j}F~.\eqno(3.2)$$

The independent components of the TM-I can be chosen as follows:
$$\eqalign{
{\rm dim-0:} ~&~ B~,\quad B^*~,\quad F~,\quad G~,\qquad\qquad
\quad ~~(4_{\rm B}) \cr
 {~~~}{\rm dim -}~\frac{1}{2}: ~&~ \de_{\pm i}F = \l_{\pm i} \quad {\rm and} \quad
\bar{\l}_{\Dot{\pm} i}~, \qquad \qquad ~(8_{\rm F}) \cr
{\rm dim-1:} ~&~ \de_{\Dot{+}i}\l\low{-j} = A\low{ij} \equiv A\low{(ij)} 
+\cc\low{ij}A~,\qquad  ~~(4_{\rm B}) 
\cr } \eqno(3.3)$$
where $\{A_{(ij)}\}^*=A^{(ij)}$ and $A$ is real. It is now straightforward to 
determine the supersymmetry transformation laws for the TM-I components from 
eqs.~(2.8) and (3.1)--(3.3). We find in addition that
$$
\de_{+i}\l_{+j}=\frac{i}{2}\cc_{ij}\de_{\dpx}B^*~,\qquad \de_{-i}\l_{+j}=0~,$$
$$
\de_{\Dot{+}i}\l\low{+j}
=-\frac{i}{2}\cc\low{ij}\de\low{\dpx}F~,\qquad \de_{\Dot{-}i}
\l\low{+j}=-A\low{ji}~,$$
$$
\de_{-i}\l_{-j}=-\frac{i}{2}\cc_{ij}\de_{\DM}B^*~,\quad \de_{+i}\l_{-j}=0~,$$
$$
\de_{\Dot{-}i}\l\low{-j}=-\frac{i}{2}\cc\low{ij}\de\low{\DM}F~, \qquad 
\de_{\Dot{+}i}\l\low{-j}=A\low{ij}~,$$
$$
\de_{+k}A=\frac{1}{2}\de_{\dpx}\l_{-k} +\frac{1}{4}(S-iT)\l_{+k}~, \qquad
\de_{-k}A=\frac{1}{2}\de_{\DM}\l_{+k} +\frac{1}{4}(S+iT)\l_{-k}~,$$
$$
\de_{+k}A_{(ij)}=i\cc_{k(i}\de_{\dpx}\l_{-j)}
-\frac{i}{2}(S-iT)\cc_{k(i}\l_{+j)}~, $$
$$
\de_{-k}A_{(ij)}=-i\cc_{k(i}\de_{\DM}\l_{+j)}
-\frac{i}{2}(S+iT)\cc_{k(i}\l_{-j)}~. \eqno(3.4)$$
Together with their complex conjugates and the defining equations it completes
the list of the (4,4) local supersymmetry transformation rules for the TM-I
components.

The TM-I is not the only minimal off-shell (4,4) hypermultiplet known to exist
in two dimensions. The different minimal hypermultiplet, TM-II, can be most 
naturally introduced after noticing that the defining constraints (2.8) of the
2d, (4,4) supergravity have additional local symmetry. Namely, they are 
invariant under the (4,4) super-Weyl transformations ({\it cf.} the super-Weyl 
symmetry of the simple $(N=1,~2d)$ superfield supergravity~\cite{howe79}):
$$ \d \de_{\pm i} = \frac{1}{2} P\de_{\pm i} + L\du{i}{j}\de_{\pm j} 
\mp (\de_{\pm i}P)\cm \pm i(\de_{\pm j}P)\cy\du{i}{j}~,\eqno(3.5)$$
and
$$\d R=PR~,\qquad \d (S-iT)=P(S-iT)
% - \frac{4}{3}\de^i\low{+}\de^j\low{\Dot{-}}L_{ij}
~,\eqno(3.6)$$
where the infinitesimal (4,4) superfield parameters $P$ and $L\du{i}{j}$ 
satisfy the constraints
$$ 
\de_{\pm k}L_{ij} ~=~\pm\,\fracm{i}{2}\left( \cc_{ik}\de_{\pm j}
+\cc_{jk}\de_{\pm i}
\right)P~,$$
$$
\[ \de_{\Dot{+}}^{~~i}~,\de_+^{~~j} \] L\low{ij}~=~
\[ \de_{\Dot{-}}^{~~i},\de_{-}^{~~j} \] L\low{ij}~=~0~.\eqno(3.7)$$
In eq.~(3.7) the signs are correlated, $L_{ij}=L_{ji}$ or $L\du{i}{i}=0$, and the 
following reality conditions are imposed:
$$ (L_{ij})^*=L^{ij}~,\qquad P^*=P~.\eqno(3.8)$$

Eqs.~(3.7) and (3.8) define the {\it twisted}-II (TM-II) hypermultiplet in the 
(4,4) superspace~\cite{gk4}. The independent components of the TM-II can be 
chosen as follows:
$$\eqalign{
{\rm dim-0:} ~&~\qquad {} \qquad P~,\qquad L\du{i}{j}~,\qquad {} \qquad
{}\qquad \qquad {} \qquad {} \qquad {}\qquad ~~~~~(4_{\rm B}) \cr
{\rm dim-}~\frac{1}{2}: ~&~ \pm\,\frac{3i}{2}\de_{\pm i}P = \c_{\pm i} \quad 
{\rm and} \quad \bar{\c}_{\Dot{\pm} i}~, \qquad {} \qquad {} \qquad\qquad
\qquad ~~~~~~(8_{\rm F}) 
\cr
{\rm dim-1:} ~&~ \de_{+}^{~~i}\de_{-}^{~~j}L_{ij} = U~,\quad {\rm and}\quad
\de_{\Dot{+}}^{~~i}\de_{\Dot{-}}^{~~j}L\low{ij}=U^*~,\cr
~&~ \de_{\Dot{+}}^{~~i}\de_{-}^{~~j}L\low{ij}=M+iN~,\quad {\rm and}\quad 
\de_{+}^{~~i}\de_{\Dot{-}}^{~~j}L\low{ij}=M-iN~, ~~(4_{\rm B})
\cr } \eqno(3.9)$$
where the $M$ and $N$ fields are real. It is straightforward to determine 
the rest of the supersymmetry transformation laws for the TM-II components. 
We find
$$
\de_{+i}\c_{+j}=0~,\qquad  \de_{-i}\c_{-j}=0~,$$
$$ 
\de_{+i}\c_{-j}=\fracm{3i}{4}R^* L_{ij}-\frac{1}{2}\cc_{ij}U~,\qquad 
\de_{-i}\c_{+j}=\fracm{3i}{4}R^* L_{ij}+\frac{1}{2}\cc_{ij}U~,$$
$$
\de_{\Dot{+}i}\c\low{+j}=-\fracm{3i}{2}\de\low{\dpx}L_{ij}+\frac{3}{4}
\cc_{ij}\de\low{\dpx}P~,\qquad 
\de_{\Dot{-}i}\c\low{-j}=-\fracm{3i}{2}\de\low{\DM}L_{ij}+\frac{3}{4}
\cc_{ij}\de\low{\DM}P~,$$
$$
\de_{\Dot{-}i}\c\low{+j}=\fracm{3i}{4}(S-iT)L\low{ij}+\frac{1}{2}
\cc_{ij}(M-iN)~,$$
$$
\de_{\Dot{+}i}\c\low{-j}=-\fracm{3i}{4}(S+iT)L\low{ij}-\frac{1}{2}
\cc_{ij}(M+iN)~, \eqno(3.10a)$$
and
$$
\de_{+i}U=-\fracm{i}{2}R^*\c_{+i}~,\qquad
\de_{-i}U=\frac{i}{2}R^*\c_{-i}~,$$
$$
\de_{\Dot{+}i}U=-\fracm{3i}{2}\de_{\dpx}\c\low{-i} 
+3i(\de_{+}^{j}S)L\low{ij} +\fracm{5i}{4}(S+iT)\c\low{+i} +\fracm{3i}{4}R^*
\c_{\Dot{+}i}~,$$
$$
\de_{\Dot{-}i}U=-\fracm{3i}{2}\de_{\DM}\c\low{+i}
+3i(\de^{~~j}_-S)L_{ij} +\fracm{5i}{4}(S-iT)\c_{i-}-\fracm{3i}{4}R^*
\c_{\Dot{-}i}~,$$

$$
\de_{+i}(M+iN)= \fracm{3i}{2}\de_{\dpx}\c\low{-i}-3i(\de_+^{~~j}S)L_{ij}
-\fracm{5i}{4}R^*\c\low{+i}-\fracm{3i}{4}(S+iT)\c\low{+i}~,$$
$$
\de_{-i}(M-iN)=\fracm{3i}{2}\de_{\DM}\c\low{+i} -3i(\de_-^{~~j}S)L_{ij}
+\fracm{5i}{4}R^*\c\low{-i} -\fracm{3i}{4}(S-iT)\c\low{-i}~,$$
$$
\de_{+i}(M-iN)= -\fracm{i}{2}(S-iT)\c\low{+i}~,$$
$$
\de_{-i}(M+iN)=-\fracm{i}{2}(S+iT)\c\low{-i}~,\eqno(3.10b)$$

Together with their complex conjugates and the defining equations it completes
the list of the (4,4) local supersymmetry transformation rules for the TM-II
components.
\vglue.2in

\section{The (2,2) supergravity solution}

In this section, we briefly review some aspects of the (2,2) extended 2d  
supergravity in N=2 superspace, and its solution as presented in 
ref.~\cite{gw1}, which are going to be relevant for our (4,4) supersymmetric 
construction in the next section.

The N=2 superspace has two real bosonic coordinates $x^{\dpx}$ and $x^{\DM}$,
and two complex fermionic coordinates $\q^+$ and $\q^-$, as well as their
conjugates $\q^{\Dot +}$ and $\q^{\Dot -}$. In addition to the N=2 superspace
general coordinate transformations, the full local symmetries of the nonminimal
 (2,2) supergravity include the local Lorentz symmetry, an axial $U_{\rm A}(1)$
 and a vector $U_{\rm V}(1)$ internal symmetries.

The geometry of the (2,2) superfield supergravity is described in terms of the
covariant spinorial derivatives
$$ \de_{\pm}= E_{\pm}{}^M\pa_M + \O_{\pm}\cm + \G_{\pm}\cx 
+ \tilde{\G}_{\pm}\tilde{\cx}~, \eqno(4.1)$$
where the generators of the local Lorentz, $U_{\rm V}(1)$ and $U_{\rm A}(1)$ 
symmetries, $\cm$, $\cx$ and $\tilde{\cx}$, respectively, have been introduced.

The nonminimal (2,2) superfield supergravity is defined by the constraints 
({\it cf.} ref.~\cite{hp})
$$ \{ \de_{\pm }\, , \de_{\pm } \} ~=~ 0~,\quad
\{ \de\low{+ } \, , \de_{\Dot{+} } \} ~=~ i \de_{\dpx}~,\quad
\{ \de\low{- } \, , \de_{\Dot{-} } \} ~=~ i \de_{\DM}~,$$
$$\eqalign{
\{ \de_{+ } , \de_-  \} ~=~& - \frac{1}{2}R^*(\cm -i\cx)~,\cr
\{ \de_{+ } , \de_{\Dot -}  \} ~=~&
      - F( \cm - i\tilde{\cx})~.\cr}\eqno(4.2)$$
The {\it minimal} N=2 supergravities appear under the restriction $F=0$ or 
$R=0$ ~\cite{hp}. 

The constraints of eq.~(4.2) are invariant under the additional local Weyl 
(scale) transformations in N=2 superspace~\cite{hp,sm1},
$$\eqalign{
E_{\pm}\to e^L E_{\pm}~,\quad & \quad  
\O_{\pm}\to e^L (\O_{\pm}\pm 2E_{\pm}L)~,\cr
\G_{\pm}\to e^L(\G_{\pm}\mp i2E_{\pm}L)~,\quad  & \quad 
\tilde{\G}_{\pm}\to e^L(\tilde{\G}_{\pm}- i2E_{\pm}L)~,\cr
R^*\to e^{2L}(R^*+4\[\de_-,\de_+\]L)~,\quad & \quad 
F\to e^{2L}(F-2i\[\de\low{\Dot -}\,,\de\low{+}\]L)~,\cr}\eqno(4.3)$$
where the Weyl superfield parameter $L$ can be restricted to be real (its
imaginary part can be absorbed by the local $U_{\rm V}(1)$ transformations). 

To solve the constraints (4.2), Grisaru and Wehlau \cite{gw1} first removed
many irrelevant superfields by imposing the supersymmetric gauge 
$$ E_+=e^{S^*}\left(\hat{E}_+ + A_+{}^-\hat{E}_-\right)~, \qquad
 E_-=e^{S^*}\left(\hat{E}_- + A_-{}^+\hat{E}_+\right)~,\eqno(4.4)$$
which does not introduce propagating ghosts. In eq.~(4.4), the reduced 
differential operators
$$ \hat{E}_{\pm}=e^{-iH^m\pa_m}D_{\pm}e^{+iH^m\pa_m}\equiv 
D_{\pm} +iH_{\pm}^m\pa_m~, \eqno(4.5)$$
a real vector superfield $H^m$ and a complex scalar superfield $S$ have been 
introduced. Substituting eq.~(4.4) into the constraints (4.2), one finds that the 
superfield connections $A_+{}^-$ and $A_-{}^+$ satisfy the {\it algebraic} 
(quadratic) equations, which determine them as the functions of $H^m$. 
The two remaining independent superfields $H^m$ and $S$ are just the (2,2) 
prepotentials of the non-minimal theory. In particular, the superfield $S$ can 
be recognized as the N=2 scale compensator since the N=2 Weyl transformation is
equivalent to a shift in $S$~\cite{gw1}. In the minimal versions of the (2,2) 
supergravity, the scale compensator is either a chiral or a twisted chiral N=2 
scalar superfield~\cite{glo,gw1,hp,sm1}. It should be noticed that the 
supersymmetric gauge-choice in eq.~(4.4) is {\it not} symmetric with respect 
to an exchange of $(-)$ and $(\Dot{-})$ objects~\cite{gw2}, so that one should
 not expect that the two minimal versions appear on equal footing from the 
non-miminal theory. 
\vglue.2in

\section{Towards a solution to the (4,4) supergravity \\ constraints}

The natural (4,4) supersymmetric generalisation of the flat (2,2) 
superdifferential operators in eq.~(4.5) is given by
$$ \hat{E}_{\pm i}=e^{-\ch}D_{\pm i}e^{\ch}~,\eqno(5.1)$$
where the operator $\ch$ of eq.~(2.5) can be restricted to have only the
`space-time' imaginary  part, $\ch=iH^{m}\pa_{m}$, with a real vector 
superfield $H^{m}$, by making certain supersymmetric gauge choices which 
do not lead to propagating ghosts. The operators $\hat{E}\low{\pm i}$, their 
conjugates $\hat{E}_{\Dot{\pm} i}$, and $\hat{E}\low{\dpx,\DM}$ to be defined below
(see eq.~(5.3)) form a convenient linearly independent basis of derivative 
operators, and satisfy a closed algebra in superspace,
$$\eqalign{
\{ \hat{E}\low{+ i}, \hat{E}_{\Dot{-}}^{~~j} \} = &
\hat{G}\du{+i\Dot{-}}{~j \dpx} \hat{E}\low{\dpx}+
\hat{G}\du{+i\Dot{-}}{~j\DM}\hat{E}\low{\DM}~,\cr
\{ \hat{E}\low{- i}, \hat{E}_{\Dot{+}}^{~~j} \} = &
\hat{G}\du{-i\Dot{+}}{~j \dpx} \hat{E}\low{\dpx}+
\hat{G}\du{-i\Dot{+}}{~j \DM}\hat{E}\low{\DM}~, \cr} \eqno(5.2)
$$
where we have introduced
$$ \hat{E}\low{\dpx}\equiv \frac{i}{2}
\{ \hat{E}\low{+}^{~i},\hat{E}_{\Dot{+} i} \}~,\quad {\rm and} \quad
\hat{E}\low{\DM} \equiv \frac{i}{2}
\{ \hat{E}\low{-}^{~i},\hat{E}_{\Dot{-} i} \}~.\eqno(5.3) $$
The `structure constants' $\hat{G}$'s in eq.~(5.2) are actually certain functions
of $\ch$ (or $H^m$, after gauge-fixing), whose expicit form is determined by 
eq.~(5.1). The full supervielbein operators should be related to that of 
eq.~(5.1), in accordance with the Frobenius theorem~\cite{book1,book2}, as
$$\eqalign{
E_{+ i}~=~&~(K_1)\du{i}{j}\left[ \hat{E}_{+ j} + A\du{+ j}{- l}\hat{E}_{- l}
\right]~,\cr
E_{- i}~=~&~(K_2)\du{i}{j}\left[ \hat{E}_{- j} + A\du{- j}{+ l}\hat{E}_{+ l}
\right]~, \cr} \eqno(5.4)$$
where the scalar $SU(2)$-tensor superfields $(K_{1,2})\du{i}{j}$ and
the vector $SU(2)$-tensor superfields $(A_a)\du{i}{j}=(A\du{+i+}{j},
A\du{-i-}{j})$ have been introduced. We have in fact assumed in eq.~(5.4) that 
the generalised `holomorphicity' takes place which allows only `undotted' indices
to appear, like in the (2,2) case. The equations for the spinorial supervielbein
operators with `dotted' indices are formally obtained from eq.~(5.4) by complex 
conjugation.

We now want to make use of the already established fact (sect.~3) that the 
two-dimensional (4,4) superspace supergravity defined by the constraints (2.8) is
superconformally flat, similarly to the $N=1$ and $N=2$ superspace supergravities
in two dimensions~\cite{howe79,hp}. It implies that the relation between the flat
and curved spinorial derivatives, as written in eq.~(5.4), should take the form 
of an (4,4) superconformal transformation. In sect.~3 we found the infinitesimal 
form of the (4,4) super-Weyl transformation but, in order to specify the matrices
$K_{1,2}$ in eq.~(5.4), we need its finite form. As  regards the (4,4) super-Weyl 
transformation law for the spinorial supervielbein componentns, one easily finds 
that
$$ (K_1)\du{i}{j}=(K_2)\du{i}{j}\equiv K\du{i}{j} = 
\exp (P{\bf 1} + L)\du{i}{j}~,\eqno(5.5)$$
where the $P$ and $L\du{i}{j}$ superfield parameters (forming a TM-II) have been
introduced in eqs.~(3.7) and (3.8), and  ${\bf 1}$ is a unit matrix.

It is straightforward to substitute our Ansatz (5.2) and (5.3) into the
constraints (2.8). As a result, all the superconnections in eq.~(2.4), as well as 
the newly introduced superfields $A$'s, are unambigously determined, as we are 
now going to demonstrate. 

First, using the constraint $\{ \de_{+ i} , \de_{+ j} \} ~=~ 0~$, we 
find 
\begin{eqnarray}
E_{+ i} E_{+ j} + \frac{1}{2} \O_{+ i} E_{+ j}
+ i~ \G\du{+ i~k}{~l} \left( \d_j^{~k} E_{+ l} - \frac{1}{2}  
\d_l^{~k}  E_{+ j}\right)~+~ ( i~ \leftrightarrow ~ j ) &=&0~,
 \nonumber \\
E_{+ i} \O_{+ j} 
+ i~ \G\du{+ i~k}{~l} \left( \d_j^{~k} \O_{+ l} - \frac{1}{2}  
\d_l^{~k}  \O_{+ j}\right)~+~ ( i~ \leftrightarrow ~ j ) &=&0~,
 \nonumber\\
i E_{+ i} \G_{+ j} \cy + \frac{i}{2} \O_{+ i} \G_{+ j} \cy 
-  \G\du{+ i~k}{~l}  \left( \d_j^{~k} \G_{+ l} \cy - \frac{1}{2}  
\d_l^{~k}  \G_{+ j} \cy \right) - 
 & & \nonumber\\
-  \G\du{+ i~r}{~l}  \G\du{+ j~l}{~s} \cy_s^{~r} 
~+~ ( i~ \leftrightarrow ~ j ) &=&0~.\nonumber
\end{eqnarray}
Since $\hat{E}_{\pm i}$ are linearly independent, it yields
$$
\frac{1}{2}(\O_{+i}\d_j^{~~l} + 2i~\G\du{+ i~j}{~l})K\du{l}{m}
~+~ ( i~ \leftrightarrow ~ j )
=-K\du{i}{k}(\hat{E}_{+k} K\du{j}{m}
+A\du{+ k}{- l} \hat{E}_{- l} K\du{j}{m})~+~ ( i~ \leftrightarrow ~ j )~,
\eqno(5.6) $$
while $A\du{+ i}{- j}$ must satisfy a differential equation
$$
\hat{E}_{+ k} A\du{+ m}{- n}+ A\du{+ k}{- l}(\hat{E}_{- l} A\du{+ m}{- n}) =0~.
\eqno(5.7) $$

By using the equation  $\{ \de_{- i} , \de_{- j} \} ~=~ 0~$, we similarly get 
\begin{eqnarray}
E_{- i} E_{- j} - \frac{1}{2} \O_{- i} E_{- j}
+ i~ \G\du{- i~k}{~l} \left( \d_j^{~k} E_{- l} - \frac{1}{2}  
\d_l^{~k}  E_{- j}\right)~+~ ( i~ \leftrightarrow ~ j ) &=&0~,
 \nonumber\\
E_{- i} \O_{- j} 
+ i~ \G\du{- i~k}{~l} \left( \d_j^{~k} \O_{- l} - \frac{1}{2}  
\d_l^{~k}  \O_{- j}\right)~+~ ( i~ \leftrightarrow ~ j ) &=&0~,
 \nonumber\\
i E_{- i} \G_{- j} \cy - \frac{i}{2} \O_{- i} \G_{- j} \cy 
-  \G\du{- i~k}{~l}  \left( \d_j^{~k} \G_{- l} \cy - \frac{1}{2}  
\d_l^{~k}  \G_{- j} \cy \right) - 
 & & \nonumber\\
-  \G\du{- i~r}{~l}  \G\du{- j~l}{~s} \cy_s^{~r} 
~+~ ( i~ \leftrightarrow ~ j ) &=&0~. \nonumber
\end{eqnarray}
It implies
$$
\frac{1}{2}(\O_{- i}\d_j^{~~l} - 2i~\G\du{- i~j}{~l})K\du{l}{m}
~+~ ( i~ \leftrightarrow ~ j )
= K\du{i}{k}(\hat{E}_{- k}K\du{j}{m}
+A\du{- k}{+ l} \hat{E}_{+ l}K\du{j}{m})~+~ ( i~ \leftrightarrow ~ j )~,
\eqno(5.8)$$
while $A\du{- i}{+ j}$ must satisfy an equation
$$
\hat{E}_{- k} A\du{- m}{+ n}+ A\du{- k}{+ l}(\hat{E}_{+  l} A\du{- m}{+ n})=0~.
\eqno(5.9) $$

The next constraint $ \{ \de_{+ i} , \de_-^{~~j}  \} ~=~ 
- \frac{i}{2}~{\Bar R}~
       \left( \d_i^{~j}\cm - \cy_i^{~j} \right)   $ yields
\begin{eqnarray}
\{ E_{+ i}, E_{- j} \} - \frac{1}{2} \O_{+ i} E_{- j} 
 + \frac{1}{2} \O_{- j} E_{+ i} +
 & & \nonumber\\
+ i~ \G\du{+ i~k}{~l} \left( \d_j^{~k} E_{- l} - \frac{1}{2}  
\d_l^{~k}  E_{- j}\right)
+ i~ \G\du{- j~k}{~l} \left( \d_i^{~k} E_{+ l} - \frac{1}{2}  
\d_l^{~k}  E_{+ i}\right) &=&0~,
 \nonumber\\
E_{+ i} \O_{- j} + E_{- j} \O_{+ i} - \O_{+ i} \O_{- j}  +
 & & \nonumber\\
+ i~ \G\du{+ i~k}{~l} \left( \d_j^{~k} \O_{- l} - \frac{1}{2}  
\d_l^{~k}  \O_{- j}\right)
+ i~ \G\du{- j~k}{~l} \left( \d_i^{~k} \O_{+ l} - \frac{1}{2}  
\d_l^{~k}  \O_{+ i}\right) &=&- \frac{i}{2} {\Bar{R}} \cc_{i j}~,
 \nonumber\\
i E_{+ i} \G_{- j} \cy + i E_{- j} \G_{+ i} \cy
- \frac{i}{2} \O_{+ i} \G_{- j} \cy 
+ \frac{i}{2} \O_{- j} \G_{+ i} \cy - 
 & & \nonumber\\
-  \G\du{+ i~k}{~l}  \left( \d_j^{~k} \G_{- l} \cy - \frac{1}{2}  
\d_l^{~k}  \G_{- j} \cy \right)
-  \G\du{- j~k}{~l}  \left( \d_i^{~k} \G_{+ l} \cy - \frac{1}{2}  
\d_l^{~k}  \G_{+ i} \cy \right) -
 & & \nonumber\\
-  \G\du{+ i~r}{~l}  \G\du{- j~l}{~s} \cy_s^{~r} 
-  \G\du{- j~r}{~l}  \G\du{+ i~l}{~s} \cy_s^{~r} 
&=&\frac{i}{2} {\Bar{R}} \cy_{i j}~. \nonumber
\end{eqnarray}
Using eqs. ~(5.6) and (5.9), we find after some algebra that
$$ \eqalign{
\O_{+i} = +K\du{i}{k}(\hat{E}_{-n}A\du{+ k}{- n}
-  A\du{+ k}{- m}\hat{E}_{+ r}A\du{- m}{+ r})~,\cr
\O_{-i} = -K\du{i}{k}(\hat{E}_{+n}A\du{- k}{+ n}
- A\du{- k}{+ n}\hat{E}_{- r}A\du{+ n}{- r})~. \cr}
\eqno(5.10) $$
We are now in a position to calculate the connections $\G\du{\pm i~j}{~l}$~.
In the matrix form, they are given by
$$ \eqalign{
\G_{+i} = \frac{i}{2}~\O_{+ i} + \frac{i}{2}E_{+ i}P 
+ i e^L U_{+i}e^{-L}~,\cr
\G_{- i}
        = \frac{i}{2}~\O_{-i}  - \frac{i}{2} E_{-i}P
- i e^L U_{+i}e^{-L}~,\cr}
\eqno(5.11) $$
where we have defined 
$$\eqalign{
U\du{+ i~j}{~k}= (e^{\frac{1}{2}P +L})\du{j}{l}(\hat{U}_{+ i}{}\du{l}{k}
+ A\du{+ l}{- m}\hat{U}\du{- i~m}{~k})~,\cr
U\du{- i~j}{~k}= (e^{\frac{1}{2}P +L})\du{j}{l}(\hat{U}_{- i}{}\du{l}{k}
+ A\du{- l}{+ m}\hat{U}\du{+ i~m}{~k})~.\cr}
\eqno(5.12) $$
To get eq.~(5.11), we used the identity
$$ \hat{E}_{\pm i}(e^L)\du{j}{k} =(e^L)\du{j}{r}(\hat{U}_{\pm i})\du{r}{k}~.
\eqno(5.13)$$
and the matrix relation
$$ \hat{E}_{\pm i}e^{\frac{1}{2}P+L}
= \frac{1}{2}(\hat{E}_{\pm i}P)e^{\frac{1}{2}P+L}
   + e^{\frac{1}{2}P} e^L \hat{U}_{\pm i}~. \eqno(5.14) $$

The constraint $
\{ \de\low{+ i} , \de_{\Dot -}^{~~j}  \} ~=~  
      - \frac{i}{2}~ ( S - i T )~ \left( \d_i^{~j}\cm - \cy_i^{~j} \right) $  
is equivalent to
\begin{eqnarray}
\{ E\low{+ i}, E_{\Dot -}^{~~j} \} - \frac{1}{2} \O\low{+ i} E_{\Dot -}^{~~j} 
 + \frac{1}{2} \O_{\Dot -}^{~~j} E\low{+ i} -
& & \nonumber\\
- i~ \G\du{+ i~k}{~l} \left( \d_l^{~j} E_{\Dot -}^{~~k} - \frac{1}{2}  
\d_l^{~k}  E_{\Dot -}^{~~j}\right) 
- i~ \G_{{\Dot -}~~k}^{~~j~~l} \left( \d_i^{~k} E_{+ l} - \frac{1}{2}  
\d_l^{~k}  E_{+ i}\right) =0~,
 \nonumber\\
E\low{+ i} \O_{\Dot -}^{~~j} + E_{\Dot -}^{~~j} \O\low{+ i} 
- \O\low{+ i} \O_{\Dot -}^{~~j}  - 
& & \nonumber\\
- i~ \G\du{+ i~k}{~l} \left( \d_l^{~j} \O_{\Dot -}^{~~k} - \frac{1}{2}  
\d_l^{~k}  \O_{\Dot -}^{~~j}\right) 
- i~ \G_{{\Dot -}~~k}^{~~j~~l} \left( \d_i^{~k} \O_{+ l} - \frac{1}{2}  
\d_l^{~k}  \O_{+ i}\right) =-\frac{i}{2} ( S - i T ) \d_i^{~j}~,
 \nonumber\\
i E\low{+ i} \G_{\Dot -}^{~~j} \cy - i E_{\Dot -}^{~~j} \G\low{+ i} \cy
- \frac{i}{2} \O\low{+ i} \G_{\Dot -}^{~~j} \cy 
- \frac{i}{2} \O_{\Dot -}^{~~j} \G\low{+ i} \cy +
 & & \nonumber\\
+  \G\low{+ i~k}{~l}  \left( \d_l^{~j} \G_{\Dot -}^{~~k} \cy - \frac{1}{2}  
\d_l^{~k}  \G_{\Dot -}^{~~j} \cy \right)
-  \G_{{\Dot -}~~k}^{~~j~~l}  \left( \d_i^{~k} \G_{+ l} \cy - \frac{1}{2}  
\d_l^{~k}  \G_{+ i} \cy \right) -
 & & \nonumber\\
-  \G\du{+ i~r}{~l} \G_{{\Dot -}~~l}^{~~j~~s} \cy_s^{~r} 
-  \G_{{\Dot -}~~r}^{~~j~~l} \G\du{+ i~l}{~s} \cy_s^{~r} 
=-\frac{i}{2} ( S - i T ) \cy_i^{~j}~. \nonumber
\end{eqnarray}
These equations yield
$$ \eqalign{
\frac{1}{2}(\O_{\Dot{-}}^{~~j}\d_i^{~l} - 2i \G_{\Dot{-}~~i}^{~~j~~l})
K\du{l}{k}~=-K\du{m}{k}(\hat{E}_{\Dot -}^{~j} K\du{i}{k}
+A\du{\Dot{-}n}{\Dot{+}m} \hat{E}_{\Dot +}^{~n}K\du{i}{k})~,\cr
\frac{1}{2}(\O_{+ i}\d_k^{~j} + 2i \G\du{+i~k}{~j})
K\du{m}{k}~=-K\du{i}{k}(\hat{E}_{+ k}K\du{m}{j}
+A\du{+ k}{- l} \hat{E}_{-l}K\du{m}{j})~,\cr} \eqno(5.15)$$
and
$$
\{ \hat{E}\low{+ k},\hat{E}_{ \Dot{-}}^{~~m} \}
- A_{\Dot{-}~~~~n}^{~~m \Dot{+}}\{ \hat{E}\low{+ k},\hat{E}_{\Dot{+}}^{~~n} \}
+ A\du{+ k}{- l}\{ \hat{E}\low{- l},\hat{E}_{ \Dot{-}}^{~~m} \}
- A\du{+ k}{- l} A_{\Dot{-}~~~~n}^{~~m \Dot{+}}
\{ \hat{E}\low{- l},\hat{E}_{\Dot{+}}^{~~n} \}=0~,
\eqno(5.16) $$
and
$$ \eqalign{ 
 \hat{E}\low{+ k}A_{\Dot{-}~~~~n}^{~~m \Dot{+}}
+ A\du{+ k}{- l}\hat{E}\low{- l} A_{\Dot{-}~~~~n}^{~~m \Dot{+}}= 0~, \cr
\hat{E}_{\Dot{-}}^{~~m} A\du{+ k}{- l} 
 -A_{\Dot{-}~~~~n}^{~~m \Dot{+}}\hat{E}_{\Dot{+}}^{~n} A\du{+ k}{- l} = 0~.\cr}
\eqno(5.17) $$

The next constraint $
\{ \de\low{+ i} , \de_{\Dot +}^{~~j}  \} ~=~  i \d_i^{~j}\de_{\dpx} $  
is equivalent to 
\begin{eqnarray}
\{ E\low{+ i}, E_{\Dot +}^{~~j} \} + \frac{1}{2} \O\low{+ i} E_{\Dot +}^{~~j} 
 + \frac{1}{2} \O_{\Dot +}^{~~j} E\low{+ i} +
& & \nonumber\\
+ i~ \G\du{+ i~k}{~l} \left(- \d_l^{~j} E_{\Dot +}^{~~k} + \frac{1}{2}  
\d_l^{~k}  E_{\Dot -}^{~~j}\right) 
- i~ \G_{{\Dot +}~~k}^{~~j~~l} \left( \d_i^{~k} E_{+ l} - \frac{1}{2}  
\d_l^{~k}  E_{+ i}\right) &=& i \d_i^{~j} \de_{\dpx},
 \nonumber\\
E\low{+ i} \O_{\Dot +}^{~~j} + E_{\Dot +}^{~~j} \O\low{+ i} -
& & \nonumber\\
- i~ \G\du{+ i~k}{~l} \left( \d_l^{~j} \O_{\Dot -}^{~~k} - \frac{1}{2}  
\d_l^{~k}  \O_{\Dot +}^{~~j}\right) 
- i~ \G_{{\Dot +}~~k}^{~~j~~l} \left( \d_i^{~k} \O_{+ l} - \frac{1}{2}  
\d_l^{~k}  \O_{+ i}\right) &=& 0 ~,
 \nonumber\\
-i E\low{+ i} \G_{\Dot +}^{~~j} \cy + i E_{\Dot +}^{~~j} \G\low{+ i} \cy
- \frac{i}{2} \O\low{+ i} \G_{\Dot +}^{~~j} \cy 
+ \frac{i}{2} \O_{\Dot +}^{~~j} \G\low{+ i} \cy -
 & & \nonumber\\
-  \G\du{+ i~k}{~l}  \left( \d_l^{~j} \G_{\Dot +}^{~~k} \cy - \frac{1}{2}  
\d_l^{~k}  \G_{\Dot +}^{~~j} \cy \right)
+  \G_{{\Dot +}~~k}^{~~j~~l}  \left( \d_i^{~k} \G_{+ l} \cy - \frac{1}{2}  
\d_l^{~k}  \G_{+ i} \cy \right) +
 & & \nonumber\\
+  \G\du{+ i~r}{~l} \G_{{\Dot +}~~l}^{~~j~~s} \cy_s^{~r} 
+  \G_{{\Dot +}~~r}^{~~j~~l} \G\du{+ i~l}{~s} \cy_s^{~r} 
&=& 0~. \nonumber
\end{eqnarray}
In particular, when $i=j$, the equations above determine $\de_{\dpx}$. When 
$i\neq j$, we find the additional constraints:
$$
\{ \hat{E}\low{+ k}, \hat{E}_{\Dot{+}}^{~~m}\}- A_{\Dot{+}~~~~n}^{~~m \Dot{-}}
\{ \hat{E}\low{+ k}, \hat{E}_{\Dot{-}}^{~~n} \}
+ A\du{+k}{-l}\{\hat{E}\low{-l},\hat{E}_{\Dot{+}}^{~m} \}
-A\du{+ k}{-l}A_{\Dot{+}~~~n}^{~~m\Dot{-}}\{\hat{E}\low{- l},
\hat{E}_{\Dot{-}}^{~~n} \} = 0~, \eqno(5.18)$$
and
$$ \eqalign{
\hat{E}\low{+k}A_{\Dot{+}~~~~~n}^{~~m\Dot{-}} + A\du{+ k}{-l}\hat{E}\low{- l}
A_{\Dot{+}~~~~~n}^{~~m\Dot{-}} = 0~, \cr
\hat{E}_{\Dot{+}}^{~~m}A\du{+ k}{-l}-
A_{\Dot{+}~~~~~n}^{~~m\Dot{-}}\hat{E}_{\Dot{-}}^{~~n}A\du{+ k}{-l} = 0~. \cr}
\eqno(5.19)$$

Finally, from the last constraint in eq.~(2.8),~ 
$\{ \de\low{- i} , \de_{\Dot -}^{~~j}  \} ~=~  i \d_i^{~j}\de_{\DM} $~,  
we find
\begin{eqnarray}
\{ E\low{- i}, E_{\Dot -}^{~~j} \} - \frac{1}{2} \O\low{- i} E_{\Dot -}^{~~j} 
 - \frac{1}{2} \O_{\Dot -}^{~~j} E\low{- i} +
& & \nonumber\\
+ i~ \G\du{- i~k}{~l} \left(- \d_l^{~j} E_{\Dot -}^{~~k} + \frac{1}{2}  
\d_l^{~k}  E_{\Dot -}^{~~j}\right) 
- i~ \G_{{\Dot -}~~k}^{~~j~~l} \left( \d_i^{~k} E_{- l} - \frac{1}{2}  
\d_l^{~k}  E_{- i}\right) &=& i \d_i^{~j} \de_{\DM}~,
 \nonumber\\
E\low{- i} \O_{\Dot -}^{~~j} + E_{\Dot -}^{~~j} \O\low{- i} -
& & \nonumber\\
- i~ \G\du{- i~k}{~l} \left( \d_l^{~j} \O_{\Dot -}^{~~k} - \frac{1}{2}  
\d_l^{~k}  \O_{\Dot -}^{~~j}\right) 
- i~ \G_{{\Dot -}~~k}^{~~j~~l} \left( \d_i^{~k} \O_{- l} - \frac{1}{2}  
\d_l^{~k}  \O_{- i}\right) &=& 0 ~,
 \nonumber\\
-i E\low{- i} \G_{\Dot -}^{~~j} \cy + i E_{\Dot -}^{~~j} \G\low{- i} \cy
+ \frac{i}{2} \O\low{- i} \G_{\Dot -}^{~~j} \cy 
- \frac{i}{2} \O_{\Dot -}^{~~j} \G\low{-i} \cy +
 & & \nonumber\\
+ \G\du{- i~k}{~l}  \left(- \d_l^{~j} \G_{\Dot +}^{~~k} \cy + \frac{1}{2}  
\d_l^{~k}  \G_{\Dot +}^{~~j} \cy \right)
+  \G_{{\Dot -}~~k}^{~~j~~l}  \left( \d_i^{~k} \G_{+ l} \cy - \frac{1}{2} 
\d_l^{~k}  \G_{+ i} \cy \right) +
 & & \nonumber\\
+  \G\du{- i~r}{~l} \G_{{\Dot -}~~l}^{~~j~~s} \cy_s^{~r} 
+  \G_{{\Dot -}~~r}^{~~j~~l} \G\du{- i~l}{~s} \cy_s^{~r} 
&=& 0~.\nonumber
\end{eqnarray}
When $i =j$, it determines $\de_{\DM}$. When $i\neq j$, some additional 
constraints on $A$'s arise, namely,
$$
\{ \hat{E}\low{-k},\hat{E}_{\Dot{-}}^{~~m} \} -
A_{\Dot{-}~~~n}^{~~m\Dot{+}}\{\hat{E}\low{- k},\hat{E}_{\Dot{+}}^{~~n} \}
-A\du{- k}{+ l}A_{\Dot{-}~~~n}^{~~m\Dot{+}}
\{\hat{E}\low{+ l},\hat{E}_{\Dot{+}}^{~~n} \}
+A\du{- k}{+ l}\{\hat{E}\low{+ l},\hat{E}_{\Dot{+}}^{~~n} \} = 0~,
\eqno(5.20)$$
and
$$ \eqalign{
\hat{E}\low{- k}A_{\Dot{-}~~~~~n}^{~~m\Dot{+}}
+ A\du{-}{+ l}\hat{E}\low{+ l}A_{\Dot{-}~~~~~n}^{~~m\Dot{+}} = 0~, \cr
\hat{E}_{\Dot{-}}^{~~m}A\du{- k}{+ l}-
A_{\Dot{-}~~~~~n}^{~~m\Dot{+}}\hat{E}_{\Dot{+}}^{~~n}A\du{- k}{+ l} = 0~. \cr}
\eqno(5.21) $$

Putting all together allows us to determine the connections $A\du{+ i}{- j}$ 
and $A\du{- i}{+ j}$. Eq.~(5.16) actually breaks up into the two equations,
$$\eqalign{
\hat{G}_{+k\Dot{-}}^{~~~~m \dpx} - i A\du{\Dot{-} k}{\Dot{+} m}
- A\du{+ k}{- l} A_{\Dot{-}~~~~ n}^{~~m \Dot{+}}
\hat{G}_{- l\Dot{+}}^{~~~~n \dpx}
 ~=~ 0~,\cr
\hat{G}_{+k\Dot{-}}^{~~~~m \DM} + i A\du{+ k}{- m}
- A\du{+ k}{- l} A_{\Dot{-}~~~~n}^{~~m \Dot{+}}
\hat{G}_{- l \Dot{+}}^{~~~~n \DM} ~=~
0~. \cr}
\eqno(5.22) $$
Similarly,  eq.~(5.18) implies 
$$\eqalign{
i\delta _k^m +A\du{+ k}{- l}\hat{G}_{- l \Dot{+}}^{~~~~m \dpx}
-A_{\Dot{+}~~~~n}^{~~m \Dot{-}}\hat{G}_{+ k \Dot{-}}^{~~~~n \dpx}
~=~ 0~, \cr
-i A\du{+ k}{- l}A_{\Dot{+}~~~~l}^{~~m \Dot{-}}+
   A\du{+ k}{- l}\hat{G}_{- l \Dot{+}}^{~~~~m \DM}
-A_{\Dot{+}~~~~n}^{~~m \Dot{-}}\hat{G}_{+ k \Dot{-}}^{~~~~n \DM} ~=~ 0~. \cr}
\eqno(5.23) $$
Eq.~(5.20) also delivers the two equations as follows:
$$\eqalign{
i\delta _k^m +A\du{- k}{+ l}\hat{G}_{+ l \Dot{-}}^{~~~~m \DM}
-A_{\Dot{-}~~~~n}^{~~m \Dot{+}}\hat{G}_{- k \Dot{+}}^{~~~~n \DM}
~=~ 0~, \cr
-i A\du{- k}{+ l}A_{\Dot{-}~~~~l}^{~~m \Dot{+}}+
   A\du{- k}{+ l}\hat{G}_{+ l \Dot{-}}^{~~~~m \dpx}
-A_{\Dot{-}~~~~n}^{~~m \Dot{+}}\hat{G}_{- k \Dot{+}}^{~~~~n \dpx} ~=~ 0~. \cr}
\eqno(5.24) $$
The first lines of eqs.~(5.23) and (5.24) are quite remarkable, since they give
the inhomogeneous {\it first-order} relations between the `undotted' and `dotted' 
components of the $SU(2)$-tensor vector superfield $A$.

The anticommutator $\{ \nabla_{\Dot{+}i}~, \de\low{-}{}^{j} \} $ implies the
second-order relations
$$
\eqalign{
\hat{G}_{- m \Dot{+}}^{~~~~~k \dpx} +i A\du{- m}{+ k}-
A_{\Dot{+}~~~~l}^{~~k \Dot{-}}A\du{- m}{+ n}\hat{G}_{+ n \Dot{-}}^{~~~~~l \dpx} 
=0~,\cr
\hat{G}_{- m \Dot{+}}^{~~~~~k \DM} -i A_{\Dot{+}~~~~m}^{~~k \Dot{-}}-
A_{\Dot{+}~~~~l}^{~~k \Dot{-}}A\du{- m}{+ n}\hat{G}_{+ n \Dot{-}}^{~~~~~l \DM} 
= 0~.\cr}
\eqno(5.25)$$
Eqs.~(5.22) and (5.25), and eqs.~(5.23) and (5.24) as well, are related by
$$
\eqalign{
(A\du{- }{+ })_i^{~j}= (A\du{+}{- }\hat{G}_{-  \dot{+}}^{~~~~ \dpx})_i^{~l}
((\hat{G}_{+  \Dot{-}}^{~~~~ \DM})^{-1})_l^{~j}~,\cr
(A_{\Dot{+}}^{~~\Dot{-}})^j_{~i}= (A_{\Dot{-}}^{~~\Dot{+}}
\hat{G}_{- \Dot{+}}^{~~~~\DM})_i^{~k}
((\hat{G}_{+ \Dot{-}}^{~~~~\dpx})^{-1})_{k}^{~j},\cr}
\eqno(5.26)$$
respectively. Altogether, they allow us to explicitly determine the superfield
$A$. In the matrix notation it takes the simple form,
$$
A\du{+i}{-j}= \left( \sqrt{(\hat{G}_{+ \Dot{-}}^{~~~~\DM})
(\hat{G}_{- \Dot{+}}^{~~~~\dpx})^{-1}} \right)\du{i}{j} ~,\qquad
A\du{-i}{+j}= \left( \sqrt{(\hat{G}_{- \Dot{+}}^{~~~~\dpx})
(\hat{G}_{+ \Dot{-}}^{~~~~\DM})^{-1}} \right)\du{i}{j}
~,\eqno(5.27a)$$
and
$$ A\du{\Dot{+}i}{\Dot{-}j} =\left(  \sqrt{(\hat{G}_{\Dot{+} -}^{~~~~\DM})
(\hat{G}_{\Dot{-} +}^{~~~~\dpx})^{-1}} \right)\du{i}{j}~,\qquad
A\du{\Dot{-}i}{\Dot{+}j}= \left( \sqrt{(\hat{G}_{\Dot{-} +}^{~~~~\dpx})
(\hat{G}_{\Dot{+} -}^{~~~~\DM})^{-1}} \right)\du{i}{j}
~.\eqno(5.27b)$$

We conclude that substituting our Ansatz into the defining constraints (2.8) leads
to both differential {\it and} algebraic equations on the superfields $A$'s. The 
algebraic constraints fully determine that superfields and, hence, fix our Ansatz
completely. The remaining differential equations (5.7), (5.9), (5.17), (5.19) and
(5.21)~\footnote{They are not all independent, but they seem to be non-trivial, 
unlike that in the (2,2) case.} become constraints on the only remaining (4,4)
superfield $H^m$. These constraints should eliminate the redundant irreducible 
(4,4) superfields in the general and reducible (4,4) superfield $H^m$, and leave 
only that (4,4) irreducible superfield which describes the off-shell (4,4) 
conformal supergravity multiplet. It is presently unclear to us how to find an 
explicit solution to the remaining highly complicated non-linear differential 
equations on the superfield $H^m$ in terms of proper (4,4) superfield 
prepotentials, beyond the linearised solution~\cite{gs,ket,kr} and a perturbation
theory.
\vglue.2in 

\section{On the matter couplings in (4,4) supergravity}

To describe the most general matter couplings in 2d, (4,4) supergravity, one
needs to describe first all inequivalent (4,4) matter representations in two 
dimensions. Constructing the most general hypermultiplet couplings in (4,4) 
supergravity remains an unsolved problem, and we are not going to solve it here. 
Instead, we want to concentrate on the (4,4) supersymmetric matter to be 
represented by TM-I or TM-II whose selfinteractions and couplings to the (4,4) 
supergravity can be rather easily constructed in superspace, like that in four 
dimensions~\cite{wpp,wpro,sam}.~\footnote{See ref.~\cite{vanpro} for a recent
review.}

The 2d, manifestly locally (4,4) supersymmetric action, which is quadratic in the
TM-I matter superfields, can be written down in terms of the TM-I prepotentials 
$(V,V\du{i}{j})$ as follows~\cite{glo}:
$$ I = \int d^2z d^8\q\,E^{-1}\left[ VA + V\du{i}{j}A\du{j}{i}\right]~,
\eqno(6.1)$$
where the full supervielbein superdeterminant $E^{-1}$ and the TM-I superfields,
$A$ and $A\du{j}{i}=(\s_I)\du{j}{i}A_I$, whose leading components are
the TM-I auxiliary fields (see sect.~3), have been introduced. The action (6.1) 
is invariant under the following gauge transformations of the 
prepotentials~\cite{sie}:
$$ \eqalign{
\d V =~&~ \de^{i\a}\L_{i\a} + {\rm h.c. }~,\cr
\d V\du{i}{j}=~&~\frac{i}{3}(\g_3)\du{\a}{\b}\de^k_{\b}\left[ \d^j_k\L^{\a}_i
-\frac{1}{2}\d^j_i\L^{\a}_k +\cc^{jl}\L^{\a}_{(ikl)}\right]~.\cr}\eqno(6.2)$$

The action (6.1) can be rewritten in the chiral superspace as
$$ I = \ha\int d^2z d^4\q\,\ce^{-1} \F^2  + {\rm h.c.}~,\eqno(6.3)$$
where the chiral superspace density $\ce$ and the reduced covariantly chiral (4,4)
superfield $\F$ (see appendix C) have been introduced. Eq.~(6.3) can be
further generalised to
$$I_V = \ha\int d^2zd^4\q\,\ce^{-1} V(\F) + {\rm h.c.}~,\eqno(6.4)$$
where $V(\F)$ is a {\it homogeneous} function of degree {\it two}, while 
maintaining all of the (4,4) superconformal symmetries. Eq.~(6.4) is quite 
similar to the standard couplings of the N=2 vector multiplets to the N=2 
supergravity in {\it four} dimensions~\cite{wpro}. However, there are also 
some important differences which originate from dimensional reduction 
(see also Appendix C).

The geometrical meaning of the action (6.4) can be most easily understood 
after rewriting it in terms of the conventional (2,2) superfields in two 
dimensions. The resulting action appears to be the special case of the general 
N=2 supersymmetric {\it non-linear sigma-model} (NLSM) coupled to the N=2 
supergravity, and described by the action
$$ I_{N=2}=\int d^2x\left\{ \int d^2\q d^2\bar{\q}E^{-1}
K(\f,\bar{\f}) - \int d^2\q \ce^{-1} W(\f) - \int d^2\q \ce^{-1}R \U(\f) 
+{\rm h.c.}\right\}~,\eqno(6.5)$$ 
in terms of the (2,2) K\"ahler potential $K$, the superpotential $W$ and 
the dilaton field $\U$ (all are functions of (2,2) chiral superfields $\f^a$
representing (2,2) matter), where $\ce$ is the (2,2) chiral density and $R$ is 
the (2,2) chiral superfield strength of the (2,2) supergravity which was already 
introduced in eq.~(4.2). In the (2,2) case, all the functions $K$, $W$ and $\U$ 
are independent off-shell, while the $W$ and $\U$ are holomorphic. On-shell, 
after eliminating the N=2 matter auxiliary fields via their algebraic equations 
of motion, that functions turn out to be related as~\cite{abk,ke5}
$$ W=\left[ \fracmm{\pa^2 K}{\pa\f^a\pa\bar{\f}^b}\right]^{-1}
\fracmm{\pa\U^*}{\pa\bar{\f}^b}\left( \fracmm{\pa W}{\pa\f^a}
+H\fracmm{\pa\U}{\pa\f^a}\right)~,\eqno(6.6)$$
where the non-propagating complex auxiliary field $H=\left.R\right|$
of the (2,2) supergravity multiplet has been introduced. In the particular case 
of a {\it single} chiral superfield $\f$, it is always possible to make the 
dilaton field linear, $\U=\f$, by field redefinition. Then eq.~(6.6) forces the 
K\"ahler metric to be flat, $K=\bar{\f}\f$, and gives rise to the {\it Liouville}
potential, $W(\f)=\m e^{\f}+H$~\cite{abk,ke5}. 

Being rewritten in the N=2 superspace to eq.~(6.5), the (4,4) supersymmetric 
action (6.4) determines all the functions  $K$, $W$ and $\U$ in terms of the only
holomorphic (and homogeneous of degree two) function $V$. One may wonder about 
the appearance of the potential and the Fradkin-Tseytlin-type term in the action 
(6.5) resulting from the action (6.4), because these terms seem to be 
inconsistent with the classical N=4 superconformal invariance of the theory under
consideration. It is nevertheless possible to have both such terms in the 
superconformally-invariant action if they appear as the result of {\it 
spontaneous} supersymmetry breaking, which triggers the spontaneous conformal 
symmetry breraking as well. Needless to say, it always leads to very special 
potentials generalising the Liouville one.

To give a simple example, let us temporarily switch off the (4,4) supergravity
fields in the (4,4) supersymmetric action (6.4). By using the results of Appendix
C and eliminating the (4,4) matter auxiliary fields via their algebraic equations
of motion, one arrives at the following bosonic part of the lagrangian describing
 the purely matter part of eq.~(6.4) ({\it cf.} ref.~\cite{kt}):
$$\eqalign{
L_{\rm B}=~&~\fracmm{\pa^2 H}{\pa N^a\pa N^b}\left\{ \pa_{\m}N^a\pa^{\m}N^b
+ \pa_{\m}M^a\pa^{\m}M^b+\pa_{\m}Q^a\pa^{\m}Q^b+\pa_{\m}P^a\pa^{\m}P^b\right\}\cr
&~ +2\fracmm{\pa^2 H}{\pa N^a\pa M^b}\ve^{\m\n}\pa_{\m}Q^a\pa^{\n}P^b
-2\fracmm{\pa^2 H}{\pa N^a\pa N^b}m^am^b \cr
&~-2m^a\fracmm{\pa^2 H}{\pa M^a\pa N^b}
\left( \left[ \fracmm{\pa^2 H}{\pa N\pa N}\right]^{-1}\right)^{bc}
\fracmm{\pa^2 H}{\pa M^c\pa N^d}m^d~,\cr}\eqno(6.7)$$
where we have used the notation (see Appendix C)
$$A=\fracmm{1}{\sqrt{2}}(M+iN)~,\quad B=\fracmm{1}{\sqrt{2}}(P+iQ)~,\quad
{\rm and}\quad H(M,N)={\rm Im}\,V(A)~.\eqno(6.8)$$
The dimensionful constants $m^a$, which appear in eq.~(6.7), arise in the process
of dimensional reduction as the expectation values of some auxiliary fields.
It now becomes clear that we are dealing with the NLSM having the torsion
and the potential induced by the spontaneous supersymmetry breaking 
$(m^a\neq 0)$ via dimensional reduction. It should also be noticed 
that, due to the torsion alone, the NLSM target space geometry in eq.~(6.7) is
{\it not} quaternionic, which agrees with the general results of de Wit and van
Nieuwenhuizen~\cite{wn}. By analogy with the NLSM counterpart in four dimensions, 
describing the scalar kinetic terms resulting from a chiral integral of a 
holomorphic function of N=2 (abelian) reduced chiral superfields~\cite{wpro},
we call the  NLSM target space geometry of eq.~(6.7) {\it special}. Eq.~(6.7) 
reduces to the free form if the function $V$ is quadratic in the fields.

Once the action (6.4) is known to have a superpotential, it must also possess 
the non-vanishing Fradkin-Tseytlin-type term because of eq.~(6.6). The action 
(6.4) may be suitable for describing the {\it non-critical} (4,4) strings 
propagating in the background spaces having special geometry~\cite{ke5}. It 
should be noticed here that quantum-mechanically consistent {\it critical} N=4 
strings do not exist, even in case of a non-trivial background space~\cite{gkw}, 
but there is no problem with constructing quantum-mechanically consistent models 
of non-critical N=4 strings~\cite{kpr}. Remarkably, no additional restrictions on
the (4,4) supersymmetric NLSM geometry arise from the NLSM quantum perturbation 
theory, since any (4,4) supersymmetric NLSM has no UV divergences at 
all.~\footnote{As far as the NLSM of eq.~(6.7) is concerned, its UV finiteness
was explicitly proved in ref.~\cite{kt} \newline ${~~~~~}$ by using the (4,4) 
superfield perturbation theory in two dimensions.} 

As far as the TM-II matter theories in the curved superspace of (4,4) supergravity
are concerned, they are extremely restricted and, until recently, no such examples
were constructed. The TM-II auxiliary fields can be considered as the leading 
components of a TM-I defining the so-called {\it kinetic} (4,4) multiplet, 
like that in 4d. Therefore, TM-I and TM-II are {\it dual} to each other, though 
they are not equivalent~\cite{gk4}. There exists the locally (4,4) supersymmetric
invariant given by a product of TM-I and TM-II~\cite{ik,ikl,gi,gk4}. In the 
curved (4,4) superspace, this invariant takes the form~\cite{ke5}
$$ I_{\rm I-II} = \int d^2x d^4\q d^4\bar{\q}E^{-1} (\P S + \X T) +\left[  
\int d^2x d^4\q\,\ce^{-1} \L R + {\rm h.c.}\right]~,\eqno(6.9)$$
where the real superfield prepotentials $\P$ and $\X$, and the chiral
superfield prepotential $\L$, of the TM-II have been introduced~\cite{gk4}.

The rigidly (4,4) supersymmetric invariant describing the free TM-II action,
which is quadratic in the fields, is known~\cite{gk4}. However, its locally
(4,4) supersymmetric generalisation does not exist.~\footnote{The same is true
in four dimensions~\cite{wpp}.} When being compared to the rigid (4,4) 
supersymmetry, the allowed matter couplings in the (4,4) conformal supergravity 
are much more restricted, and it is also known to be the case for the N=2 matter 
couplings in the four-dimensional N=2 supergravity~\cite{wpro}. As far as the 
TM-II in 2d is concerned, this problem is only apparent, since there exists its 
{\it improved} (i.e. superconformally invariant) 2d action~\cite{ke5}, which can 
be coupled to the 2d, (4,4) conformal supergravity. The point is that it is 
possible to form the TM-I out of the TM-II components in yet another 
{\it non-linear} way, namely,
$$\eqalign{
R_{\rm impr.}~=~&~  L^{-1} U^*
+  \fracm{4 i}{3} \c\low{\Dot{+}i} 
{\c\low{\Dot{-}}}^{j} L\du{j}{i} L^{-3}~,\cr
S_{\rm impr.}~=~&~ \fracmm{1}{4}~ L^{-1} M
+ \fracmm{2i}{3} \left( \c\low{\Dot{+}i} {\c\low{-}^{j}} - 
{\c\low{\Dot{-}}}^j {\c\low{+i}} \right) L\du{j}{i} L^{-3}~,\cr
T_{\rm impr.}~=~&~  \fracmm{1}{4}~L^{-1} N 
+\fracmm{2}{3} \left( 
\l\low{\Dot{+}i} {\c\low{-}}^j + {\c\low{\Dot{-}}}^j {\c\low{+i}} \right)
  L\du{j}{i} L^{-3}~,\cr}\eqno(6.10)$$
where we have used the notation
$$ L\equiv \sqrt{L\du{i}{j}(L\du{i}{j})^*}~.\eqno(6.11)$$ 

As was shown in sect.~3, the (4,4) superspace constraints (2.8) have the hidden 
super-Weyl symmetry, the (4,4) supergravity field strengths are represented by 
TM-I, and TM-II appears as a scale compensator. The non-linear realisation of 
TM-I can be derived from a calculation of the {\it finite} form of the super-Weyl
transformations. One finds eq.~(6.10) either in the lowest order of an expansion 
of the finite super-Weyl transformation in powers of the TM-I fields, or, 
equivalently, in a superconformally flat gauge where the (untransformed) TM-I 
fields are set to zero. The infinitesimal super-Weyl transformations given in 
eq.~(3.6) vanish in the superconformally flat gauge.

Eq.~(6.9) can now be used to define an invariant coupling of the improved 
TM-II to the (4,4) supergravity in the form
$$ I_{\rm impr.} = \int d^2x d^4\q d^4\bar{\q}E^{-1} (\P S_{\rm impr.}
+ \X T_{\rm impr.}) +\left[  \int d^2x d^4\q\ce^{-1}
\L R_{\rm impr.} + {\rm h.c.}\right]~.\eqno(6.12)$$
The existence of the improved TM-II in 2d is a direct consequence of the 
existence of the improved N=2 tensor multiplet in 4d~\cite{wpp}, since they 
are related via dimensional reduction. Unlike the improved N=2 tensor 
multiplet in 4d, its 2d counterpart does not have any gauge degrees of 
freedom, which allows the (4,4) locally supersymmetric component action 
associated with eq.~(6.12) to have the manifest $SU(2)$ internal symmetry.

There actually exists an additional resource to build yet another (4,4) locally
supersymmetric invariant, namely, the so-called {\it Fayet-Iliopoulos} (FI) 
term.~\footnote{The FI term was used in ref.~\cite{gi} to construct the rigidly 
(4,4) supersymmetric Liouville action.} In the curved (4,4) superspace, the FI 
term takes the form~\cite{ke5}
$$ I_{\rm FI} = -2\m  \int d^2x d^4\q d^4\bar{\q}E^{-1} \P~.\eqno(6.13)$$
Given the action
$$ I_{\rm L}=I_{\rm impr.} + I_{\rm FI}~,\eqno(6.14)$$
the auxiliary field $M$ of the improved TM-II enters this action in the
combination $e^{-\f}M^2-2\m M$. Eliminating this auxiliary field via its
algebraic equation of motion gives rise to the {\it Liouville} potential 
$-\m^2e^{\f}$ again. The action (6.14) is therefore the (4,4) locally 
supersymmetric Liouville action.
\vglue.2in

\section{Conclusion}

In this paper we considered the superfield structure of the (4,4) conformal
supergravity in two dimensions, and made progress in finding a solution to
its superspace constraints. Even though the remaining problems are of technical 
nature, a deeper insight into the complicated superspace structure of the (4,4)
supergravity theory may be needed in order to get the explicit solution. As a
next task, a detailed comparison with the linearised analysis may be useful, in
order to find the best way to proceed.    

Another aspect deserving further investigations is the (4,4) locally 
supersymmetric model-building, i.e. constructing the matter couplings in 2d, (4,4)
supergravity. We discussed in this paper only two (4,4) scalar multiplets, TM-I
and TM-II, whereas different variant representations of hypermultiplet are
also known to exist~\cite{gk4}. It would be of interest to describe them also. 
The (4,4) non-critical strings and the NLSM special geometry are the natural 
areas for posssible applications of such models. 

The very framework of the conventional (4,4) superspace used above may happen to 
be inadequate for describing the {\it most} general matter couplings in the (4,4)
supergravity, so that the more powerful {\it harmonic} superspace method 
\cite{hrs} may be needed.  The $SU(2)\times SU(2)$ harmonic (4,4) superspace 
approach recently proposed by Ivanov and Sutulin~\cite{is,i} may be the proper 
way to address general issues. The $SU(2)\times SU(2)$ harmonic (4,4) superspace 
has two independent sets of harmonic variables and the necessarily infinite sets 
of auxiliary fields for an off-shell hypermultiplet and an off-shell (4,4) 
supergravity, which make the transition from any harmonic superspace formulation 
to components highly non-trivial. The exisiting resources of the conventional 
(4,4) superspace deserve to be explored further, in parallel with the 
complementary harmonic superspace approach.

We summarize the component results about the 2d, (4,4) supergravity and the (4,4)
string action in Appendix B. The list of symmetries of the (4,4) string action is
also given in Appendix B. It includes both the known continuous symmetries and 
the new discrete symmetries of the string action.
\vglue.2in

\section*{Acknowledgements}

One of the authors (S.V.K.) would like to thank Jim Gates, Marc Grisaru, Evgeni 
Ivanov, Jens Schnittger and Marcia Wehlau for many stimulating discussions. 

\newpage

{\Large\bf Appendix A: notation and conventions}
\vglue.1in

We use small greek letters $\l,\m,\n,\ldots$ for the vector indices associated
with the two-dimensional curved spacetime or the string world-sheet, and small 
latin letters $a,b,c,\ldots$ for the vector indices in the corresponding 
tangent space. The two-dimensional Minkowski metric is given by
$$ \h_{ab} = \left( \begin{array}{cc} -1  &  0 \\
 0 & 1 \end{array} \right)~.\eqno(A.1)$$

Given $\e_i$ to represent a Dirac spinor in the fundamental representation of
$SU(2)$, small latin letters $i,j,k,\ldots$ are used to denote the $SU(2)$
indices, $i=1,2$. The $SU(2)$ indices are `canonically' contracted from the upper
left to the lower right, and they are raised (lowered) with  $\cc^{ij}~ 
(\cc_{ij})$, so that the following identities hold:
$$ \e^i\equiv\cc^{ij}\e_j~,~~~\e_i\equiv\e^j\cc_{ji}~,~~~
\cc^{ik}\cc_{kj}= - \d^{i}_{~j}~.\eqno(A.2)$$
Explicitly, we have
$$ \cc^{ij} = i
 \left( \begin{array}{cc} 0  &  1 \\
 -1 & 0 \end{array} \right)~.\eqno(A.3)$$
The complex conjugation acts on the $SU(2)$ indices in the following way:
$$ (\e_i)^{*}\equiv\e^{*i}~.\eqno(A.4)$$
Thus, we deduce that $\cc_{ij}=(\cc^{ij})^{*}$ and
$$ (\e^i)^{*}=(\cc^{ik}\e_k)^{*}= \cc_{ik}\e^{*k}=-\e\ud{*}{i}~.\eqno(A.5)$$
The Majorana and Dirac conjungations of spinors are defined as
follows:~
$$\tilde{\e}^{i}=(\e^i)^{\rm T}C~,~~~~~\bar{\e}^i=(\e_i)\dg C~,\eqno(A.6)$$
where $C_{\a\b}=\s^2$ is  the charge conjugation matrix which obeys
$$ C=C\dg ~,\qquad C \g^a C^{-1}=-(g^a)^{\rm T}~.\eqno(A.7)$$

We find it useful to introduce the light-cone coordinates
$$ x^{\pm} = \fracmm{1}{\sqrt{2}} \left( x^0 \pm x^1 \right)~,\eqno(A.8)$$
in terms of the coordinates $x^a =( x^0, x^1)$ of the tangent space. The index
values $a=0,1$ here should not be confused with the similar values for the
target space indices.

The light-cone components $\e_{\pm}$ of a spinor $\e$ define the
one-dimensional representations of the Lorentz group $SO(1,1)$. The field
$\e_{+}$  $(\e_{-})$ moves to the right (left), and they have the following
transformation properties under the action of the $SO(1,1)$  generator $\g_3$:
$$ \g_3 \e_{-} =  - \e_{-}~,\qquad \g_3 \e_{+}=\e_{+}~.\eqno(A.9) $$
We identify the spinor components of $\e$ with its light-cone components 
$\e_{\pm}$, 
$$ \e_i = \left( \begin{array}{cc} {\e_{+ i}}  \\
 {\e_{- i}}  \end{array} \right)~.\eqno(A.10)$$

The two-dimensional gamma matrices satisfy the algebra
$$ (\g^a)\du{\a}{\d}(\g^b)\du{\d}{\b}=\h^{ab}\d\du{\a}{\b}
+\ve^{ab}(\g_3)\du{\a}{\b}~,\eqno(A.11)$$
where $\ve^{ab}$ is the  Levi-Civita symbol, $ \ve^{01}=1$. Explicitly, we
choose $(\g_0)\du{\a}{\b}=-i\s^2~$, $(\g_1)\du{\a}{\b}=\s^1~$, and
$(\g_3)\du{\a}{\b}=(\g^0\g^1)\du{\a}{\b}=\s^3~,$ or, equivalently,
$$ \g_0 = \left( \begin{array}{cc} 0  &  -1 \\
 1 & 0 \end{array} \right)~,\qquad
\g_1 = \left( \begin{array}{cc} 0  & 1 \\
 1 & 0 \end{array} \right)~,\qquad
\g_3 = \left( \begin{array}{cc} 1  &  0 \\
 0 & -1 \end{array} \right)~.\eqno(A.12)$$

In our component formulae to be given in Appendix B, all the spinor indices
are omitted, as a rule. In addition, the following identities hold:
$$\eqalign{
\{\g^a,\g^b\}=~&~2\h^{ab}~,\cr
\[\g^a,\g^b\]=~&~2\ve^{ab}\g_3~,\cr
\ve^{ab}\ve_{cd}=~&~-(\d_c^a\d_d^b- \d_d^a\d_c^b)~,\cr
\g_3\g_a=~&~\ve_{ab}\g^b~.\cr}\eqno(A.13)$$

As far as the curved 2d spacetime or the string world-sheet is concerned,
we have the following relation for the gamma matrices:
$$\g^\m\g^\n=\h^{\m\n} +e^{-1}\ve^{\m\n}\g_3~,\eqno(A.14)$$
where  $e$ is the determinant of the zweibein $e\du{\m}{a}$, and
$e^{-1}\ve^{\m\n}$ is the Levi-Civita tensor density.

The Pauli matrices $(\s^I)\du{i}{j}$ satisfy the algebra
$$ (\s^I~\s^J )\du{i}{j} = \d^{IJ}\d\du{i}{j}+ i\ve^{IJK}
 (\s^K)\du{i}{j}~,\qquad
 (\s^I)_{ij}\equiv (\s^I)\du{i}{k}\cc_{kj}= (\s^I)_{ji}~,\eqno(A.15)$$
and have the usual form
$$ (\s^1)\du{i}{j} = \left( \begin{array}{cc} 0  &  1 \\
 1 & 0 \end{array} \right)~,~~~
(\s^2)\du{i}{j} = \left( \begin{array}{cc} 0  & -i \\
 i & 0 \end{array} \right)~,~~~
(\s^3)\du{i}{j} = \left( \begin{array}{cc} 1  &  0 \\
 0 & -1 \end{array} \right)~.\eqno(A.16)$$

A calculation of the spinor bilinear relations in N=4 supersymmetry is similar to
that in N=1 or N=2 supersymmetry, but it also has some additional features due to
 the $SU(2)$ structure. For instance, we find that
$$\eqalign{
\tilde{\e}^i\j_i ~&~=
 (\e^i)^TC\j_i = ( \cc^{ij}\e_{j} )^T C \j_{i} =
  ( \cc^{ij} )^T \e\du{j}{\a} C_{\a\b} \j\du{i}{\b}
 =  \j\du{i}{\b} C_{\b\a} ( \cc^{ij} )^T \e^{j\a} \cr
&~ = ( \cc^{ij} \j_{i} )^T C \e_{j} = -(\j^{i} )^T C \e_{i}
 = -\tilde{\j}^i\e_i~.\cr}\eqno(A.17)$$
Eq.~(A.17) implies, in particular, that  $\tilde{\j}^i\j_i = 0$. As far as the
spinor bilinears with Pauli matrices are concerned, we find
$$\eqalign{
\tilde{\e}^i(\s^I)\du{i}{j}\j_j ~&~=
 ( \cc^{ik} \e_k )^TC(\s^I)\du{i}{j}\j_j =
 (\e_k)^TC \cc^{ki}(\s^I)\du{i}{j}\j_j =
  \e\du{k}{\a}C_{\a\b}(\s^I)^{kj} \j\du{j}{\b} \cr
 ~&~ = \j\du{j}{\b} C_{\b\a}(\s^I)^{jk}\e\du{k}{\a}
 = (\j_{j})^T C \cc^{ji}(\s^I)\du{i}{k}\e_{k}
 =   (  \cc^{ij} \j_{j})^T C (\s^I)\du{i}{k}\e_{k} \cr
~&~ = \tilde{\j}^i(\s^I)\du{i}{k}\e_k~. \cr}\eqno(A.18)$$

Because of the relation $C\g^a C^{-1}=-(g^a)^{\rm T}$, the contraction of spinor
indices over a gamma matrix yields
$$\tilde{\e}^i\g_a\j_i= \tilde{\j}^i\g_a\e_i~.\eqno(A.19)$$

\vglue.3in

{\Large\bf Appendix B: (4,4) supergravity in components}

\vglue.2in

An off-shell multiplet $(e\du{\m}{a},\j_{\m i},A\du{\m}{I},R,S,T )$ of the 2d,
minimal $N=4$ conformal supergravity was given in refs.~\cite{glo,ghn89}
(see also refs.~\cite{pn86,sch87} for the earlier results). We have
$e\du{\m}{a}$ for the zweibein, a complex  Dirac
spinor $\j_{\m i}$ in the $SU(2)$ doublet-representation for the gravitini, and
$A\du{\m}{I}$ as the real $SU(2)$ gauge field in the triplet-representation.
The scalars $S,T$ and $R$ are all the auxiliary fields. The fields $S$ and $T$
are real, whereas the field $R$ is complex. Alltogether, this gives $(8 + 8)$
bosonic and fermionic degrees of freedom off-shell.

The infinitesimal transformation laws for the 2d, N=4 conformal supergravity
fields read ({\it cf.\/} refs.~\cite{glo,ghn89})
$$ \eqalign{
\d_Q e\du{\m}{a} = & -\ha\bar{\e}^i\g^a\j_{\m i}
 +\ha\bar{\j}\du{\m}{i} \g^a\e_i~,\cr
\d_Q \j_{\m i} = & D_\m\e_i+\g_\m\left[ \frac{1}{4}(S-i\g_3T)\e_i
+\frac{i}{2}R\g_3\e\ud{*}{i}\right]~, \cr
\d_Q A\du{\m}{I} = & \frac{i}{4}\bar{\e}^i
(\s^I)\du{i}{j}\g_\m\g_3\ve^{\r\s} D_\r\j_{\s j}
-\frac{i}{4}\bar{\e}^i(\s^I)\du{i}{j}\g_\m\g^\r
\left[ \frac{1}{4}(S-i\g_3T)\j_{\r j} +\frac{i}{2}R\g_3\j\ud{*}{\r j}\right]
\cr
 & -\frac{i}{4}\bar{\j}\du{\n}{i}\g_\m\g^\n(\s^I) \du{i}{j}\left[
\frac{1}{4}(S-i\g_3T)\e_j +\frac{i}{2}R\g_3\e\ud{*}{j}\right] + {\rm h.c.}~,
\cr
\d_Q S = & -\ve^{\m\n}\bar{\e}^i\g_3D_\m\j_{\n i}
+\bar{\e}^i\g^\m
\left[ \frac{1}{4}(S-i\g_3T)\j_{\m i}
+\frac{i}{2}R\g_3\j\ud{*}{\m i}
\right] + {\rm h.c.}~,\cr
\d_Q T = & i\ve^{\m\n}\bar{\e}^iD_\m\j_{\n i}
-i\bar{\e}^i\g_3\g^\m
\left[ \frac{1}{4}(S-i\g_3T)\j_{\m i}
+\frac{i}{2}R\g_3\j\ud{*}{\m i}
\right] + {\rm h.c.}~,\cr
\d_Q R = & i\ve^{\m\n}\tilde{\e}^iD_\m\j_{\n i}
-i\tilde{\e}^i\g_3\g^\m
\left[ \frac{1}{4}(S-i\g_3T)\j_{\m i}
+\frac{i}{2}R\g_3\j\ud{*}{\m i}
\right] ~, \cr
\d_Q R^{*}= & -i\ve^{\m\n}\bar{\e}^iD_\m\j\ud{*}{\n i}
+i\bar{\e}^i\g_3\g^\m
\left[ \frac{1}{4}(S+i\g_3T)\j\ud{*}{\m i}
+\frac{i}{2}R^{*}\g_3\j_{\m i} \right] ~,\cr}\eqno(B.1)$$
where we have introduced $D_\m$ as the covariant derivative,
$$ D_\m \e_i=\pa_{\m} \e_i +\frac{1}{4}\o\du{\m}{a b}(e,\j)\ve_{a b}\g_3 \e_i
-iA\du{\m}{I}(\s_I)\du{i}{j}\e_j~.\eqno(B.2)    $$
The spin connection $\o\du{\m}{a b}(e,\j)$ is given by
$$\eqalign{
\o\du{\m}{a b}(e,\j) = & \o\du{\m}{a b}(e)-\frac{1}{4} \left(\right.
\bar{\j}\du{\m}{i}\g^a\j\ud{b}{i} -\bar{\j}\du{\m}{i}\g^b\j\ud{a}{i}\cr
 & + \bar{\j}^{a i}\g_\m\j\ud{b}{i}
-\bar{\j}^{b i}\g^a\j_{\m i}+\bar{\j}^{a i}\g^b\j_{\m i}
-\bar{\j}^{b i}\g_\m\j\ud{a}{i} \left.\right) ~, \cr}\eqno(B.3)$$
where $\o\du{\m}{a b}(e)$ is the usual (torsion-free) spin connection,
$$\eqalign{
\o\du{\m}{a b}(e) = &
\ha e^{\n a}(~ \pa_\m e\du{\n}{b}-\pa_\n e\du{\m}{b}~ )
 -\ha e^{\n b}(~ \pa_\m e\du{\n}{a}-\pa_\n e\du{\m}{a}~ )\cr
 & -\ha e^{\r a} e^{\l b}(~ \pa_\r e_{\l c}-\pa_\l e_{\r c}~ ) e\du{\m}{c}~.
\cr}\eqno(B.4)$$

\vglue.2in

{\Large\bf Appendix C: $N=2$ real chiral superfield in four dimensions, and its
dimensional reduction to $d=2$ }
\vglue.1in

It is often useful to formulate field theories with extended supersymmetry in
higher dimensions and then dimensionally reduce them to lower dimensions. As far
as the 2d, $N=4$ field theories are concerned, one can use
either the $N=1$ superfields in six dimensions or the $N=2$ superfields in four
dimensions (4d). Taking the latter choice, the simplest $N=2$ superspace 
constraints defining an $N=2$ chiral scalar superfield $\F (x^{\m},\q^{\a}_i,
\bar{\q}^{\dt{\a}i})$, $\m=0,1,2,3$ and $i,j,\ldots =1,2$, are given by~\footnote{
 In this Appendix C we use the standard four-dimensional notation~\cite{bw,kr}.}
$$ D^i_{\a}\Bar{\F}=\Bar{D}_{\dt{\a}i}\F=0~,\eqno(C.1)$$
where we have introduced the (flat) superspace covariant derivatives
$ D^i_{\a}$ and $\Bar{D}_{\dt{\a}i}$ satisfying the algebra
$$\left\{ D^i_{\a},\Bar{D}_{j\dt{\b}}\right\}=2i\d^i_j\s^{\m}_{\a\dt{\b}}
\pa_{\m}~.\eqno(C.2)$$
The complex supermultiplet $(\F,\Bar{\F})$ is reducible to the (generalised) 
real one by using the additional constraint~\cite{w}
$$\fracmm{1}{12} D^{\a}_iD_{\a j}D^{\b i}D^j_{\b}\F=\Box \Bar{\F}~.\eqno(C.3)$$
The solution to the constraints (C.2) and (C.3) reads~\footnote{We define 
$(\q^3)^{i\a}=(\pa/\pa\q_{i\a})\q^4$ and
$\q^4=\fracmm{1}{12}\q^{\a}_i \q_{\a j}\q^{\b i}\q^j_{\b}$.}
$$\eqalign{
\F =~&~ \exp\left\{ -\fracmm{i}{2}\q^{\g}_i\tilde{\s}^{\m}_{\g\dt{\g}}\pa_{\m}
\bar{\q}^{\dt{\g}i}\right\}\left[ A + \q^{\a}_i\j^i_{\a} -
\ha\q^{\a}_i(\s^I)\ud{i}{j}\q^j_{\a}C^I \right. \cr
&~\left. +
\fracmm{1}{8}\q^{\a}_i(\s_{\m\n})\du{\a}{\b}\q^i_{\b}F^{\m\n} -
i(\q^3)^{i\a}\tilde{\s}^{\m}_{\a\dt{\b}}\pa_{\m}\Bar{\j}^{\dt{\b}}_i+
\q^4\Box \Bar{A} \right]~,\cr}\eqno(C.4)$$
in terms of the components 
$$ \left( A, \j^i_{\a},C^I,F_{\m\n}\right)~,\eqno(C.5)$$
where $A$ is a complex scalar, $\j^i$ is a 4d Majorana spinor isodoublet, $C^I$
is a real isovector, and $F_{\m\n}$ is a real antisymmetric tensor satisfying
the constraint 
$$ \pa^{\m}F_{\m\n}=0~.\eqno(C.6)$$
Because of the constraint (C.6), the tensor $\tilde{F}_{\m\n}$ dual to 
$F_{\m\n}$ can be interpreted as the field strength of a vector~\cite{w,ga}. 
Accordingly, the supermultiplet (C.5) is usually referred to as the $N=2$ 
{\it vector} multiplet in four dimensions.

The dimensional reduction to two dimensions amounts to $\pa_2=\pa_3=0$. The 
4d isospinor $\j^i$ can then be represented in terms of 2d spinors as
$$\j^i=\left(\begin{array}{c} \j^i \\ \tilde{\j}^i \end{array}\right)~,\qquad
 \tilde{\j}^i \equiv \cc^{ij}C\bar{\j}^{\rm ~T}_j~,\eqno(C.7)$$
where $C$ is the 2d charge conjugation matrix defined in Appendix A.

The constraint (C.6) can be easily  solved {\it after} the dimensional 
reduction~\cite{kt},
$$\eqalign{
F_{01}=m={\rm const}~,\quad & \quad F_{23}=D~,\cr
F_{\m 2}=\frac{1}{2}\ve_{\m\n}\pa^{\n}(B+\Bar{B})~,\quad & \quad 
F_{\m 3}=\frac{1}{2i}\ve_{\m\n}\pa^{\n}(B-\Bar{B})~,\cr}\eqno(C.8)$$
in terms of a complex scalar $B$ and a real scalar $D$, where an arbitrary
dimensionful constant $m$ appears, in general. As a result, one arrives
at the 2d, $N=4$ twisted scalar multiplet,
$$\left( A,B,\j^i,C^I,D\right)~,\eqno(C.9)$$
comprising two complex scalars $A$ and $B$, a 2d Dirac spinor isodoublet
$\j^i$ and the auxiliary fields: a real isovector $C^I$ and a real scalar 
$D$ ($8_{\rm B}\oplus 8_{\rm F}$ components). It is called the TM-I, according to
the classification proposed in ref.~\cite{gk4}. The transformation laws for the
TM-I version of hypermultiplet components, which are obtained via dimensional 
reduction from the transformation laws of the 4d, N=2 vector multiplet are given 
by~\cite{kt}
$$\eqalign{
\d A =~&~ \bar{\ve}_i\frac{1}{2}(1-\g_3)\j^i +
\bar{\j}^i\frac{1}{2}(1-\g_3)\ve_i~,\cr
\d B =~&~ \bar{\ve}_i\g_3\tilde{\j}^i~,\cr
\d\j^i=~&~ (\s^I)\ud{i}{j}C^I\g_3\ve^j -\frac{1}{2}(1-\g_3)i\tilde{\pa}A\ve^i
 +\frac{1}{2}(1+\g_3)i\tilde{\pa}\Bar{A}\ve^i \cr
~&~ -iD\ve^i + 2m\g_3\ve^i +i\tilde{\pa}\tilde{\ve}^i\Bar{B}~,\cr
\d C^I=~&~ -\frac{1}{2}\bar{\ve}_i\tilde{\pa}(\s^I)\du{j}{i}\j^j 
+{\rm h.c.}~,\cr
\d D =~&~  \frac{1}{2}\bar{\ve}_i\g_3\tilde{\pa}\j^i +{\rm h.c.}~,\cr}
\eqno(C.10)$$
where $\ve^i$ are the 2d, infinitesimal (4,4) supersymmetry parameters  forming 
a Dirac spinor isodoublet, $\tilde{\pa}=\g^{\m}\pa_{\m}$, and $\g^{\m}$, $\m=0,1$,
 are the 2d Dirac matrices defined in Appendix A.

The non-vanishing dimensionful constant $m$ triggers a spontaneous breakdown of 
the (4,4) supersymmetry. It is already ovbious from the (4,4) supersymmetry 
transformation law for the spinor fields $\j^i$ in eq.~(C.10) whose right-hand
side contains the Goldstone term (see sect.~6 also).
\vglue.2in

\end{document}